\documentclass[prl,twocolumn,superscriptaddress,amsmath,amssymb]{revtex4-1}
\usepackage{graphicx,bm,dsfont}
\usepackage{epstopdf}
\usepackage[colorlinks=true,citecolor=blue,urlcolor=blue]{hyperref}

\begin{document}
\title{Dissipative Topological Phase Transition with Strong System-Environment Coupling}

\author{Wei Nie}
\affiliation{Theoretical Quantum Physics Laboratory, RIKEN Cluster for Pioneering Research, Wako-shi, Saitama 351-0198, Japan}
\author{Mauro Antezza}
\affiliation{Laboratoire Charles Coulomb (L2C), UMR 5221 CNRS-Universit\'{e} de Montpellier, F- 34095 Montpellier, France}
\affiliation{Institut Universitaire de France, 1 rue Descartes, F-75231 Paris Cedex 05, France}
\author{Yu-xi Liu}\email{yuxiliu@mail.tsinghua.edu.cn}
\affiliation{School of Integrated Circuits, Tsinghua University, Beijing 100084, China}
\affiliation{Frontier Science Center for Quantum Information, Beijing, China}
\author{Franco Nori}\email{fnori@riken.jp}
\affiliation{Theoretical Quantum Physics Laboratory, RIKEN Cluster for Pioneering Research, Wako-shi, Saitama 351-0198, Japan}
\affiliation{RIKEN Center for Quantum Computing (RQC), 2-1 Hirosawa, Wako-shi, Saitama 351-0198, Japan}
\affiliation{Physics Department, The University of Michigan, Ann Arbor, Michigan 48109-1040, USA}

\begin{abstract}
A primary motivation for studying topological matter regards the protection of topological order from its environment. In this work, we study a topological emitter array coupled to an electromagnetic environment. The photon-emitter coupling produces nonlocal interactions between emitters. Using periodic boundary conditions for all ranges of environment-induced interactions, the chiral symmetry inherent to the emitter array is preserved. This chiral symmetry protects the Hamiltonian, and induces parity in the Lindblad operator. A topological phase transition occurs at a critical photon-emitter coupling related to the energy spectrum width of the emitter array. Interestingly, the critical point nontrivially changes the dissipation rates of edge states, yielding a dissipative topological phase transition. In the protected topological phase, edge states suffer from environment-induced dissipation for weak photon-emitter coupling. However, strong coupling leads to robust dissipationless edge states with a window at the emitter spacing. Our work shows the potential to manipulate topological quantum matter with electromagnetic environments.
\end{abstract}

\maketitle
\textit{Introduction}.---Vacuum electromagnetic environments can nontrivially change order parameters of matter, producing phase transitions~\cite{landig2016quantum,PhysRevX.10.041027}. With the advances in cavity quantum electrodynamics (QED)~\cite{Haroche2006book,gu2017microwave,kockum2019ultrastrong,mivehvar2021cavity}, vacuum electromagnetic fields are used to manipulate matter~\cite{PhysRevLett.122.133602,PhysRevLett.122.167002,thomas2019exploring,Garcia2021} with strong light-matter interaction. For example, in cavity-interfaced superconductors, a strong coupling with electromagnetic fields changes the superconducting transition temperature~\cite{thomas2019exploring}. Recently, the vacuum electromagnetic control of matter is receiving growing attention~\cite{gonzalez2015subwavelength,PhysRevLett.124.083603,rui2020subradiant}. Due to symmetry-protected properties, topological matter is also being studied in the coupling with electromagnetic fields for potential applications~\cite{PhysRevLett.122.236803,PhysRevResearch.2.012076,mann2020topological}. The bandgap of a kagome metasurface of dipole emitters embedded in a cavity can be tuned by electromagnetic fields~\cite{mann2020tunable}. Varying the cavity width can change long-range interactions between emitters and induce topological phase transitions~\cite{mann2020topological}.

A prerequisite to make topological protection reliable is to understand dissipative properties of topological systems~\cite{diehl2011topology,PhysRevB.84.205109,PhysRevB.85.121405,PhysRevB.88.205142,bardyn2013topology,shen2014hall,PhysRevB.92.165118,PhysRevB.94.201105,
PhysRevX.7.041062,carollo2018uhlmann,PhysRevB.99.125118,weisbrich2019decoherence,carollo2020geometry,PhysRevB.102.165116,gneiting2020unraveling,Leefmans2021}. Energy bands play a pivotal role for topological matter, e.g., in studying topological phases~\cite{atala2013direct,zhang2019topological,RevModPhys.91.015005,RevModPhys.91.015006} and topological criticalities~\cite{PhysRevB.93.165423,PhysRevE.96.020106,PhysRevB.102.134213}. The large gap between energy bands protects topological properties from local disorder~\cite{bravyi2010topological,hafezi2011robust,PhysRevB.85.165124,PhysRevLett.113.046802,poli2015selective,PhysRevB.96.035149,PhysRevLett.122.076801,PhysRevLett.122.193903,PhysRevLett.124.023603} and thermal noises~\cite{PhysRevLett.109.105302,PhysRevLett.112.130401,PhysRevLett.113.076407,PhysRevLett.117.226801,PhysRevX.8.011035,PhysRevLett.125.215701}. However, a recent study~\cite{McGinley2020fragility} of time-reversal symmetry protected topological systems with large bandgap shows the fragility of topological phases in electromagnetic environments. Via \emph{perturbation} theory, they find that quantum coherence between edge states in one-dimensional (1D) topological systems is spoiled when system-environment coupling is weak compared to the bandgap. This finding shows the challenge of protecting topological quantum matter in electromagnetic environments.

\begin{figure}[t]
\includegraphics[width=8.5cm]{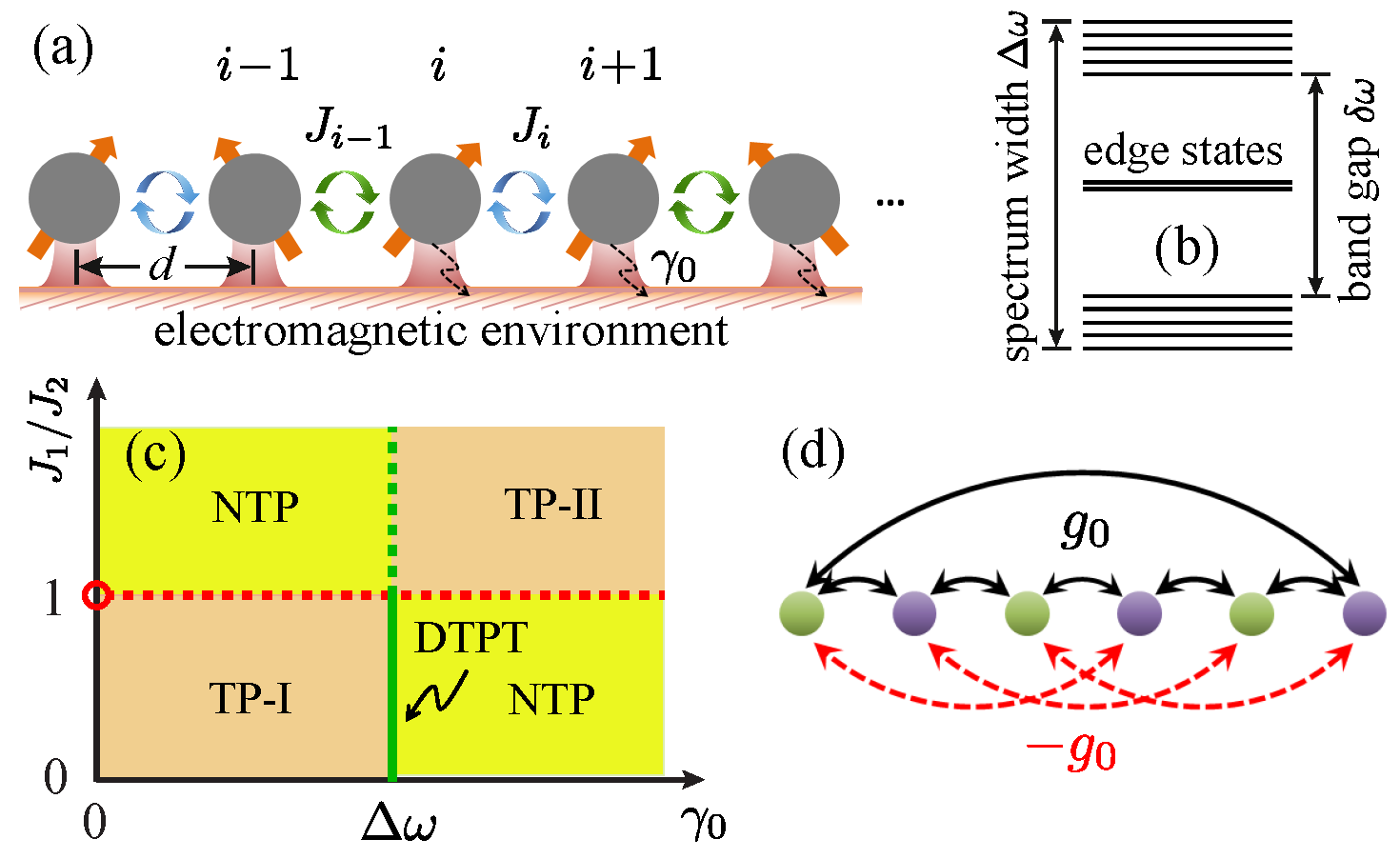}
\caption{(a) Schematic of a topological emitter array coupled to an electromagnetic environment. Emitters have nearest-neighboring interactions $J_i$, homogeneous spacing $d$, and decay rate $\gamma_0$ to the environment. (b) Single-excitation spectrum of the topological system with spectrum width $\Delta\omega$ and bandgap $\delta\omega$. (c) Phase diagram of the system for emitter spacing $d=3\lambda_0/4$. There are topological phases TP-I (having edge states), TP-II (having no edge state) and non-topological phase (NTP). The red circle at ($\gamma_0=0$, $J_1/J_2=1$) represents the original phase transition in the SSH model. The horizontal red-dashed line denotes the SSH-type phase transition in the dissipative regime. The vertical green line represents the environment-induced topological phase transition where the decay rate $\gamma_0$ is equal to the spectrum width $\Delta\omega$ of the topological system. In particular, the green-solid line indicates the dissipative topological phase transition. (d) Photon-mediated interactions $H_{\mathrm{ph}}$ for $d=\lambda_0/4$ ($g_0=\gamma_0/2$) and $d=3\lambda_0/4$ ($g_0=-\gamma_0/2$).}\label{fig1}
\end{figure}

In this work, we study the coupling between a topological emitter array and its electromagnetic environment in the \emph{nonperturbative} regime, i.e., edge states are coupled to bulk states via the environment. We find that for emitter spacings $d=\lambda_0/4$ and $d=3\lambda_0/4$, environment-induced interactions  have chiral symmetry and produce distinct topological phases. For $d=\lambda_0/4$, environment modifies the topological phase with dissipative edge states. However, the edge states for $d=3\lambda_0/4$ are protected from dissipation in a parameter space specified by the Lindblad operator. In the thermodynamic limit, a dissipative topological phase transition (DTPT), characterized by a nontrivial change of dissipation of the edge states, occurs at $d=3\lambda_0/4$ when the single-emitter decay rate induced by the system-environment coupling equals the energy spectrum width of the topological emitter array. These results could be useful for improving topological protection in open quantum systems which have nonlocal dissipations~\cite{McGinley2020fragility}.

\textit{1D topological emitter array in vacuum electromagnetic fields}.---We consider a topological array of dipole emitters coupled to its surrounding electromagnetic environment, as in Fig.~\ref{fig1}(a). The single-excitation energy spectrum of the topological emitter array with bandgap and spectrum width is shown in Fig.~\ref{fig1}(b). Electromagnetic modes in the environment are described by
$H_{\mathrm{E}}=\int d^3\bm{r}\int_{0}^{\infty} d\omega\;\hbar \omega\;\hat{a}^{\dagger}(\bm{r,\omega})\hat{a}(\bm{r},\omega)$,
where $\hat{a}^{\dagger}(\bm{r,\omega})$ and $\hat{a}(\bm{r},\omega)$ are the creation and annihilation operators of photons. The emitter-environment coupling is
$H_{\mathrm{int}}=-\sum_i \int_0^{\infty} d\omega(\hat{\bm{d}}_i \cdot \bm{E}(\bm{r}_i,\omega)+ \mathrm{H.c.})$, where $\hat{\bm{d}}_i= \bm{d}_i \sigma_i^- + \bm{d}_i^{\ast}\sigma_i^+$ is the dipole moment operator of the $i$th emitter. The electric field operator is
$\bm{E}(\bm{r},\omega)=i \eta \int d^3 \bm{r}' \sqrt{\varepsilon_I(\bm{r}',\omega)} \bm{G}(\bm{r},\bm{r}',\omega) \hat{a}(\bm{r}',\omega)$, where $\eta=\sqrt{\hbar}\omega^2/\sqrt{\pi \epsilon_0}c^2$; $\varepsilon_I(\bm{r}',\omega)$ is the imaginary part of the complex permittivity; the Green's tensor $\bm{G}(\bm{r},\bm{r}',\omega)$ describes the electromagnetic interaction from $\bm{r}'$ to $\bm{r}$. The dynamics of the topological emitter array is described by the master equation~\cite{breuer2002theory,SupplementalMaterial}
\begin{equation}
\dot{\rho}(t)=-\frac{i}{\hbar}[H_0 + H_{\mathrm{topo}}+H_{\mathrm{ph}},\rho(t)] + \mathcal{D}[\rho], \label{Eqmaster}
\end{equation}
where the free energy is $H_0=\sum_i \hbar\omega_0 \sigma_i^+\sigma_i^-$ ($\omega_0$ is the transition frequency of emitters) and the topological emitter array is described by $H_{\mathrm{topo}}=\sum_i \hbar J_i(\sigma_i^+ \sigma_{i+1}^- + \sigma_{i+1}^+\sigma_i^-)$ with tunable dimerized interactions $J_{i}=J_0[1+(-1)^i\cos\varphi]$~\cite{PhysRevLett.113.220502}. The emitter-environment coupling can be strong compared with the bandgap, but much smaller than the energy of emitters, to satisfy the Born-Markov approximation in Eq.~(\ref{Eqmaster}). The virtual-photon exchange between emitters and environment yields $H_{\mathrm{ph}}=\sum_{i,j=1}^{N} \hbar g_{ij} (\sigma_{i}^- \sigma_{j}^+ + \sigma_{j}^- \sigma_{i}^+)$, where $g_{ij}$ [Eq.~(\ref{Eqg})] characterize the strengths of the nonlocal dipole-dipole interactions. In addition to the coherent part $H_{\mathrm{ph}}$, the virtual-photon exchange yields correlated dissipations $\gamma_{ij}$ [Eq.~(\ref{Eqgamma})], which are included in the Lindblad operator,
\begin{equation}
\mathcal{D}[\rho]=\sum_{i,j=1}^{N} \gamma_{ij} \left(\sigma_{i}^- \rho \sigma_{j}^+ - \frac{1}{2}\sigma_{i}^+ \sigma_{j}^- \rho - \frac{1}{2}\rho \sigma_{i}^+ \sigma_{j}^-\right). \label{Eqdissipation}
\end{equation}

By applying the Kramers-Kronig relation to the Green's tensor and integrating in the frequency domain, the photon-mediated interactions and dissipations become~\cite{knoll2000qed,PhysRevA.66.063810,PhysRevB.82.075427,PhysRevLett.106.020501,PhysRevB.84.235306,PhysRevA.91.051803,PhysRevA.95.033818,PhysRevA.95.063807,PhysRevX.7.031024,PhysRevLett.119.173901}
\begin{eqnarray}
g_{ij}&=& \frac{\omega_0^2}{\hbar \varepsilon_0 c^2} \mathrm{Re}[\bm{d}_i^\ast \cdot \bm{G}(\bm{r}_i,\bm{r}_j,\omega_0) \cdot \bm{d}_j], \label{Eqg} \\
\gamma_{ij}&=& \frac{2\omega_0^2}{\hbar \varepsilon_0 c^2} \mathrm{Im}[\bm{d}_i^\ast \cdot \bm{G}(\bm{r}_i,\bm{r}_j,\omega_0) \cdot \bm{d}_j]. \label{Eqgamma}
\end{eqnarray}
For the 1D electromagnetic environment, concrete forms of the nonlocal interactions and correlated dissipations are~\cite{PhysRevLett.106.020501,PhysRevB.84.235306,PhysRevA.92.053834,PhysRevA.93.033833,mirhosseini2019cavity,PhysRevLett.123.233602,zanner2021coherent} $g_{ij} = \gamma_0 \sin (2\pi d_{ij}/\lambda_0)/2$ and $\gamma_{ij} = \gamma_0 \cos (2\pi d_{ij}/\lambda_0)$, respectively. Here, the emitter decay rate is $\gamma_0=g^2/c$ where $g$ is the photon-emitter coupling and $c$ is the group velocity of photons; $d_{ij}$ is the distance between $i$th and $j$th emitters; $\lambda_0$ is the wavelength of a photon with frequency $\omega_0$. We find that the \emph{spectrum width} $\Delta\omega$ sets a critical point for a dissipation-induced topological phase transition, represented by the green-solid vertical line in Fig.~\ref{fig1}(c).

\textit{Environment-protected chiral symmetry}.---As a simple illustration, in Fig.~\ref{fig1}(d) the environment induces nearest-neighboring (NN) and long-range interactions in an array with $N=6$ emitters. We consider the cases when the spacings $d=\lambda_0/4$ and $d=3\lambda_0/4$; and the parameter $g_0$ is $\gamma_0/2$ and $-\gamma_0/2$, respectively. The long-range interaction between the first and the last emitters provides periodic boundary conditions for the NN interaction. Conversely, the long-range interaction between the $i$th and $(i+5)$th emitters exhibits translational invariance due to the NN interaction. Therefore, the effective strengths for the NN interaction and the long-range interaction between the $i$th and $(i+5)$th emitters, are $g_0/2$. Moreover, the effective interaction $g_{ij}$ between the $i$th and $(i+3)$th emitters (red-dashed curves) is $-g_0/2$. With this protocol, the translational symmetry is preserved for all ranges of interactions induced by the environment at $d=\lambda_0/4$ and $d=3\lambda_0/4$. However, for other values of the spacing $d$, the translational symmetry in $H_{\mathrm{ph}}$ is broken.

\begin{figure}[t]
\includegraphics[width=8.5cm]{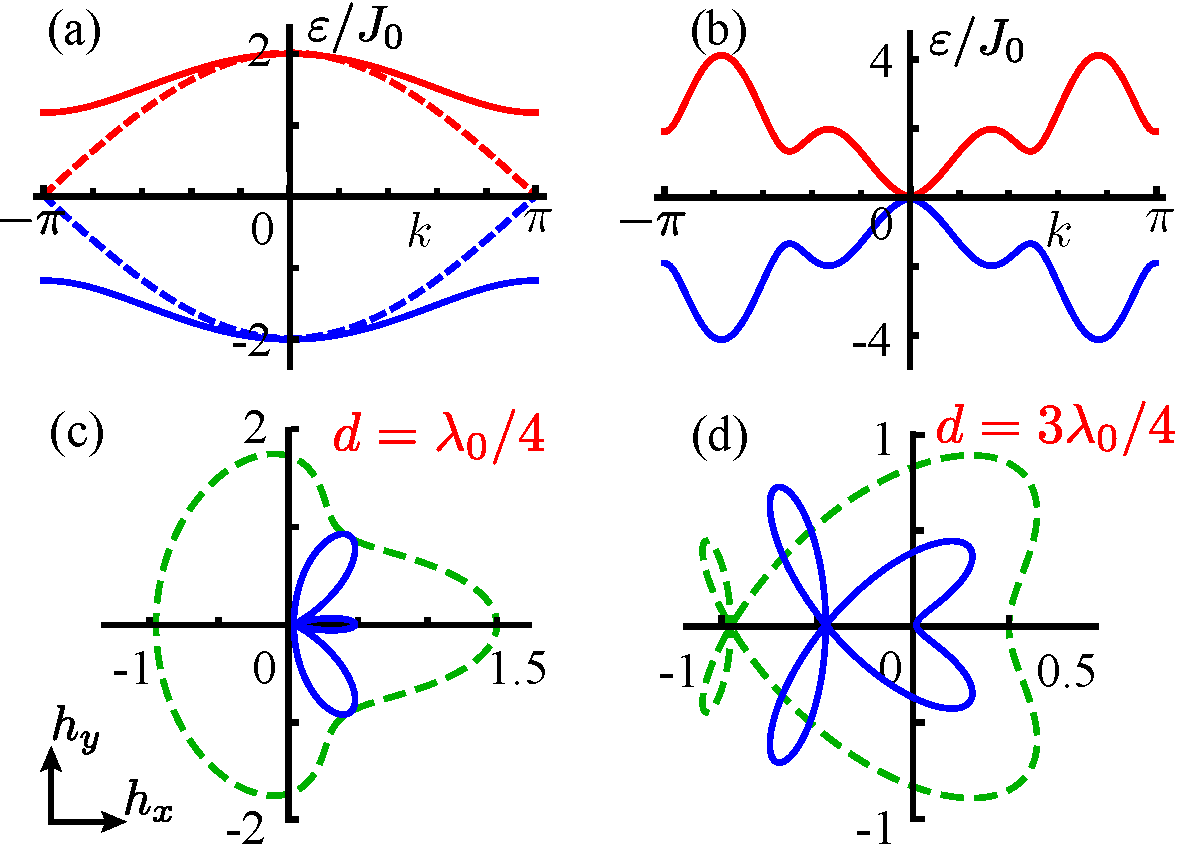}
\caption{(a) Energy bands of the topological emitter array for $J_1\neq J_2$ (solid) and $J_1=J_2$ (dashed). (b) Environment-induced gap closing at emitter spacing $d=3\lambda_0/4$ and decay rate $\gamma_0=\Delta\omega$. Topologies from the hybridization between $H_{\mathrm{topo}}$ and $H_{\mathrm{ph}}$ in auxiliary space $(h_x(k), h_y(k))$ for (c) $d=\lambda_0/4$, and (d) $d=3\lambda_0/4$. (c) The winding number is zero at $J_0=0$ (blue-solid), and becomes one for $J_0>0$ (green-dashed). (d) The winding number is zero for $0\leq J_0 \leq \gamma_0/4$ (the blue-solid topology denotes $J_0=\gamma_0/4$), and becomes one for $J_0>\gamma_0/4$ (green-dashed). We consider $\varphi=0.1\pi, N=6$.}\label{fig2}
\end{figure}

By assuming periodic boundary conditions on $H_{\mathrm{topo}}$, the coherent interaction $H=H_{\mathrm{topo}}+H_{\mathrm{ph}}$ in quasi-momentum space is $H/\hbar=\sum_k \Psi^{+}_k \mathcal{H}(k)\Psi_k$, where $\Psi^{+}_k=(\sigma_{A,k}^{+},\sigma_{B,k}^{+})$. Here, $A$ and $B$ denote odd- and even-site emitters, respectively. The 1D symmetry-protected topological system is described by the Su-Schrieffer-Heeger (SSH) model~\cite{PhysRevLett.42.1698}. In the sublattice space, we obtain an effective spin-$1/2$ Hamiltonian  $\mathcal{H}(k)= h_x(k)\tau_x + h_y(k)\tau_y$ with chiral symmetry $\tau_z\mathcal{H}(k)\tau_z=-\mathcal{H}(k)$~\cite{asboth2016short}. Here, $\tau_x, \tau_y, \tau_z$ are Pauli matrices, and
\begin{eqnarray}
h_x(k) &=& J_1 +J_2 \cos (k) + \frac{g_0}{2} \left[1+ \cos \left(\frac{Nk}{2}\right)\right], \\
h_y(k) &=& J_2 \sin (k) + \frac{g_0}{2} \mathcal{F}(k),
\end{eqnarray}
with $g_0=\gamma_0/2$ ($-\gamma_0/2$) for $d=\lambda_0/4$ ($3\lambda_0/4$), $\mathcal{F}(k)=\sum_{j=1}^{N/2} 2 (-1)^{j-1} \sin (jk) - \sin (N k/2)$, and energy bands $\varepsilon_{\pm}(k)=\pm\sqrt{h_x^2(k) + h_y^2(k)}$. Without the environment, the energy bands are shown in Fig.~\ref{fig2}(a). The bandgap and spectrum width are
\begin{equation}
\delta\omega=2|J_1-J_2|, \quad \Delta\omega=2(J_1 + J_2),
\end{equation}
respectively. The dimerized interactions $J_{1,2}=J_0(1\mp\cos\varphi)$ yield the bandgap $\delta\omega=4J_0\cos\varphi$ and spectrum width $\Delta\omega=4J_0$. The SSH-type topological phase transition takes place at $k=\pm \pi$~\cite{asboth2016short} with linear low-energy dispersion. In the electromagnetic environment with emitter spacing $d=3\lambda_0/4$, the condition
\begin{equation}
\gamma_0=\Delta\omega,
\end{equation}
yields a gap closing at $k=0$ with parabolic dispersion, as shown in Fig.~\ref{fig2}(b). The parabolic dispersion~\cite{PhysRevB.93.165423,PhysRevE.96.020106} makes this topological criticality to be different from the one in the SSH model. In the auxiliary space $(h_x(k), h_y(k))$, the winding number can be defined as $W=(1/2\pi)\int_{\mathrm{B.Z.}} d\theta_k$, with $\theta_k=\arctan [-h_x(k)/h_y(k)]$. For $d=\lambda_0/4$, shown in Fig.~\ref{fig2}(c), the system is in a non-topological phase with $W=0$ at $J_0=0$. However, as $J_0$ is increased, the winding number $W=1$; i.e., the topological phase is protected when $d=\lambda_0/4$. For $d=3\lambda_0/4$, in Fig.~\ref{fig2}(d), the system has zero winding number for small $J_0/\gamma_0$. However, at a critical point
$\gamma_0^c=\Delta\omega$, a topological phase transition takes place. For $\gamma_0<\gamma_0^c$, the system becomes topological with winding number $W=1$. Namely, the topological phase is preserved when the spectrum width $\Delta\omega$ is larger than the environment-induced decay $\gamma_0$ of the emitters, as shown in Fig.~\ref{fig1}(c).

\begin{figure}[t]
\includegraphics[width=8.5cm]{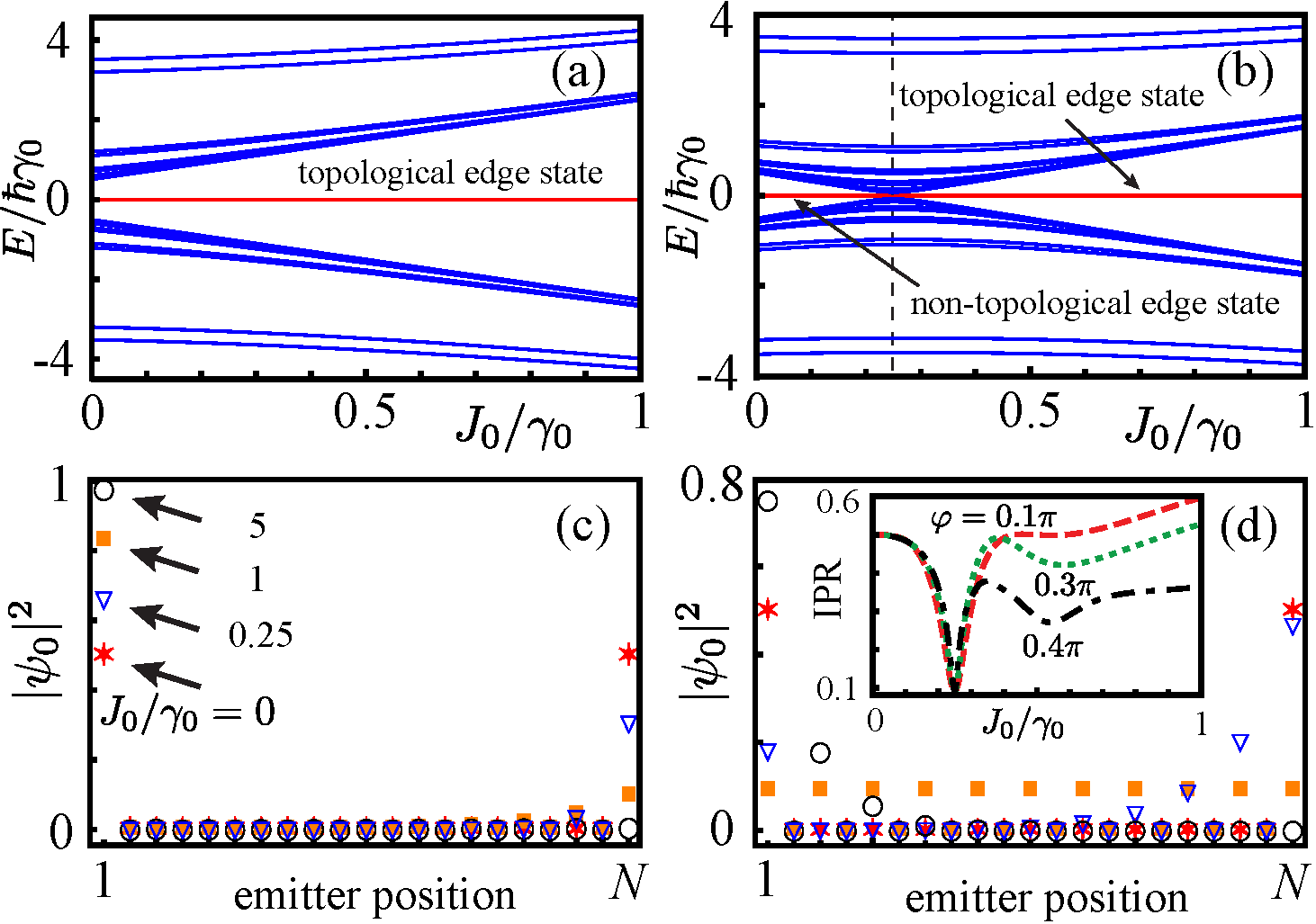}
\caption{Energy spectra for (a) $d=\lambda_0/4$, and (b) $d=3\lambda_0/4$, respectively. Probability distribution $|\psi_0|^2$ of the zero-energy state for (c) $d=\lambda_0/4$ and (d) $d=3\lambda_0/4$. In (c,d), red stars, blue triangles, orange squares, and black circles correspond to $J_0/\gamma_0=0, 0.25, 1, 5$, and $J_0/\gamma_0=0, 0.2, 0.25, 1$, respectively. The inset of (d) shows the IPR of the zero-energy state at different values of $\varphi$ for $d=3\lambda_0/4$. In (a)-(d) we consider $\varphi=0.1\pi$, $N=21$.}\label{fig3}
\end{figure}

\textit{Edge state vs dissipative topological phase transition}.---Figures~\ref{fig3}(a,b) show the energy spectra of $H$ versus $J_0/\gamma_0$ for (a) $d=\lambda_0/4$ and (b) $d=3\lambda_0/4$ in a system with an odd number of emitters $N=21$, where a single edge state appears. In agreement with the topologies in quasi-momentum space for these two emitter spacings, a bandgap [\ref{fig3}(a)] and a band touching [\ref{fig3}(b)] are found. In Fig.~\ref{fig3}(b), a non-topological edge state is found for the topologically trivial phase. Figures~\ref{fig3}(c,d) show the distributions $|\psi_0|^2$ of the edge state. At $J_0=0$, the edge state is equally distributed at the two edge emitters with wave function $|\psi_0\rangle=\frac{1}{\sqrt{2}} (\sigma_{1}^+ + \sigma_{N}^+) |G\rangle$, where $|G\rangle$ is the ground state of the emitter array. In Fig.~\ref{fig3}(c), with $d=\lambda_0/4$, enlarging $J_0$ monotonically increases the component of $|\psi_{0}|^2$ at the left boundary. However, before the critical point, the left-boundary component of the edge state for $d=3\lambda_0/4$ becomes smaller as $J_0/\gamma_0$ is increased. At the critical point, the gap of the spectrum closes and the edge state becomes delocalized. By further increasing $J_0$, the edge state eventually localizes at the left boundary.

To characterize the changes of the edge state, we study the inverse participation ratio (IPR)~\cite{PhysRevB.83.184206}, $\mathrm{IPR} = \sum_i |\psi_{0i}|^4/(\sum_i |\psi_{0i}|^2)^2$, where $\psi_{0i}$ is the amplitude of the edge state at the $i$th emitter. In the inset of Fig.~\ref{fig3}(d), we show the IPR versus $J_0/\gamma_0$ for $d=3\lambda_0/4$. The IPR of the edge state at $J_0=0$ is one half due to its equal distribution at two boundaries. A minimum is found at the critical point for different values of $\varphi$, indicating the edge-bulk transition.

\begin{figure}[t]
\includegraphics[width=8.5cm]{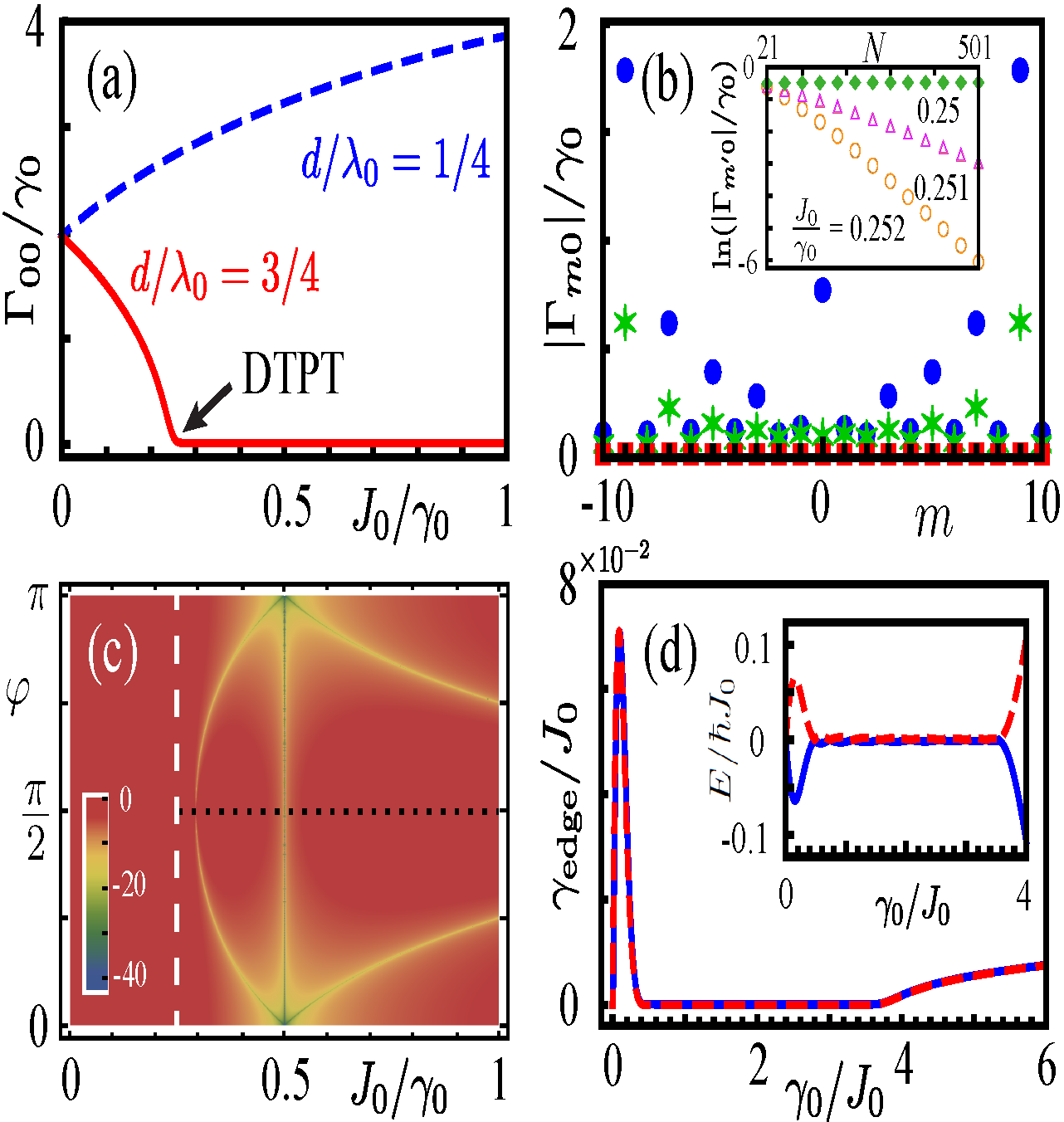}
\caption{(a) Dissipation $\Gamma_{00}$ from edge state to environment for $d=\lambda_0/4$ (blue-dashed) and $d=3\lambda_0/4$ (red-solid). (b) Dissipations $\Gamma_{m0}$ from edge state to environment ($m=0$) and bulk states ($m\neq0$) for $d=3\lambda_0/4$ at $J_0/\gamma_0=0.2$ (blue dots), $0.25$ (green stars) and $0.3$ (red squares). The inset shows  $\mathrm{ln}(|\Gamma_{m'0}|/\gamma_0)$ versus $N$ for $J_0/\gamma_0=0.25$ (green diamonds), $0.251$ (purple triangles) and $0.252$ (orange circles). (c) $\mathrm{ln}(\Gamma_{00}/\gamma_0)$ for the emitter array with $N=7$. The white-dashed vertical ($J_0/\gamma_0=1/4$) and black-dotted horizontal lines indicate the DTPT and the SSH-type criticality, respectively. (d) Dissipations from edge states to environment. The inset shows the corresponding energy levels of the edge states. We consider $N=21$ [one edge state] in (a),(b), $N=20$ [two edge states] in (d); $\varphi=0.3\pi$ in (a),(b),(d).}\label{fig4}
\end{figure}

To study the stability of topological features in real space, we here rewrite the Lindblad operator in terms of eigenstates of $H$,
\begin{equation}
\mathcal{D}[\rho]=\sum_{m,n}\Gamma_{mn}[\Psi_m^-\rho\Psi_n^+ -\frac{1}{2} \Psi_m^+\Psi_n^-\rho -\frac{1}{2} \rho \Psi_m^+\Psi_n^-], \label{EqLindblad}
\end{equation}
with $\Psi_m^+=|\Psi_m\rangle \langle G|$. Here, ${|\Psi_m\rangle}$ denotes the $m$th eigenmode of $H$. The decay rates are $\Gamma_{mn}=\sum_{i,j}\gamma_{ij}\langle e_i| \Psi_m \rangle \langle \Psi_n| e_j\rangle$, with $|e_i\rangle=\sigma_i^+ |G\rangle$. Specifically, $\Gamma_{mm}$ denotes the decay rate of the $m$th eigenstate to environment; $\Gamma_{mn}$ is the correlated decay from the $n$th state to $m$th state. The dissipation of the edge state is governed by $\Gamma_{m0}$. In Fig.~\ref{fig4}(a), we show the scaled decay rate $\Gamma_{00}/\gamma_0$ from edge state to environment versus $J_0/\gamma_0$. For $d=\lambda_0/4$, $\Gamma_{00}/\gamma_0$ increases with $J_0/\gamma_0$, and decreases after reaching the maximum~\cite{SupplementalMaterial}. However, the edge state at $d=3\lambda_0/4$ has a decay rate that decreases in the non-topological phase and that stops decaying at $J_0=\gamma_0/4$. In finite systems, the weak emitter-environment coupling, i.e., small $\gamma_0/J_0$, introduces dissipation of the edge state~\cite{SupplementalMaterial}, which is responsible for the enhanced photon absorption~\cite{nie2020topology}. However, the edge state for strong coupling is protected against decoherence in the topological phase.

For added clarity, the correlated decays $\Gamma_{m0}$ ($m\neq0$) between the edge state and the bulk states are shown in Fig.~\ref{fig4}(b). At $J_0/\gamma_0=0.2$ (blue dots) in the non-topological phase, the edge state not only decays into the environment ($\Gamma_{00}\neq0$), but also decays into the bulk states of the emitter array. However, at $J_0/\gamma_0=0.3$ (red squares) in the topological phase, the edge state does not decay to bulk states. At the critical point $J_0/\gamma_0=0.25$, the dissipations to bulk states are greatly suppressed, except for those of the two bulk states $m'=\pm(N-3)/2$. Near the critical point, the correlated dissipations
$|\Gamma_{m'0}|\propto \exp(-\nu_{m'} N)$, with $\nu_{m'}>0$ in the topological phase. The inset shows $\mathrm{ln}(|\Gamma_{m'0}|/\gamma_0)$ versus $N$. The values of $\nu_{m'}$ are $0$, $0.005$, and $0.0115$, for $J_0/\gamma_0=0.25$, $0.251$, and $0.252$, respectively. Therefore, in the thermodynamic limit $N \rightarrow +\infty$, the critical point indicates a transition between dissipative and dissipationless edge states, namely, a DTPT, which can be accessed by observing the population dynamics of the emitter array~\cite{SupplementalMaterial}. In Fig.~\ref{fig4}(c), the local minima of $\mathrm{ln}(\Gamma_{00}/\gamma_0)$ show the parameter space of edge states protected by the Lindblad operator~\cite{PhysRevX.6.041031,PhysRevA.98.042118,PhysRevA.100.062131,PhysRevLett.124.040401} in a small system. This protection is actually attributed to the vanishing overlap between edge states and polarized radiating modes in the Lindblad operator, which exhibits parity property, i.e., dissipations only occur between odd-site (even-site) emitters. Larger systems have broader parameter space for dissipationless edge states~\cite{SupplementalMaterial}. In particular, the condition that the environment-induced decay is half of the spectrum width, i.e., $J_0/\gamma_0=1/2$, produces a dissipationless edge state
\begin{equation}
\left|\psi_0\left(\frac{J_0}{\gamma_0}\rightarrow \frac{1}{2}\right)\right\rangle =\frac{1}{\sqrt{\mathcal{N}}}\sum_{n\in\mathbb{N}} (-1)^{n}\left(\tan\frac{\varphi}{2}\right)^{2n} |\psi\rangle_n,
\end{equation}
for various localization lengths even near the SSH criticality. Here, $|\psi\rangle_n=(\sigma_{4n+1}^+ + \sigma_{4n+3}^+)|G\rangle$; namely, the $(4n+1)$th and $(4n+3)$th emitters have the same amplitude.

\textit{Dissipationless subspace of topological edge states}.---In arrays with an even number of emitters, two edge states appear at the boundaries. Figure~\ref{fig4}(d) shows the decay rates and energy splitting (in the inset) of the edge states. With small $\gamma_0$ (weak coupling), the two localized edge states are coupled by environment-mediated long-range interactions, and the subspace of edge states suffers from decoherence~\cite{McGinley2020fragility,PhysRevLett.127.086801}. Surprisingly, when the emitter-environment coupling is strong, i.e., $\gamma_0$ is large, the edge states are decoupled from each other. Therefore, they are both protected from dissipation until the DTPT at $\gamma_0=\Delta\omega$. Moreover, strong coupling makes the zero splitting between edge states insensitive to emitter spacing around $d=3\lambda_0/4$~\cite{SupplementalMaterial}.

\begin{figure}[t]
\includegraphics[width=8.5cm]{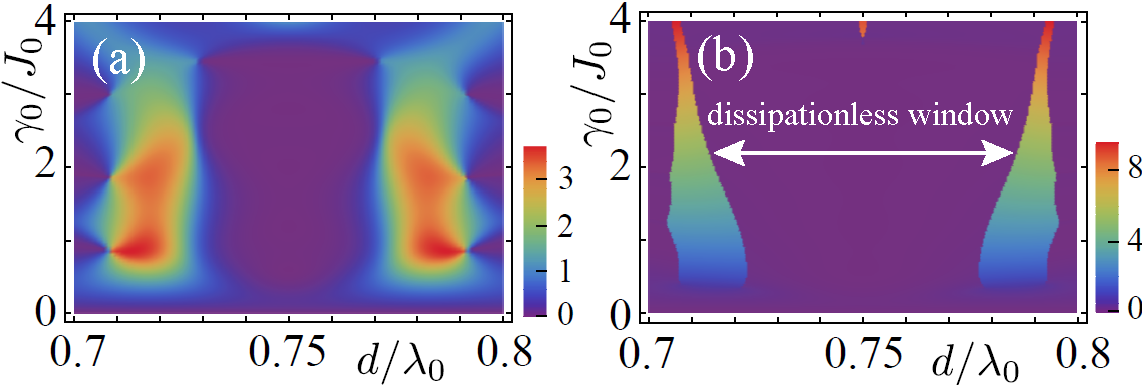}
\caption{(a) Decay rate $\Gamma_{00}$ from edge state to environment. (b) Effective decay rate $\tilde{\Gamma}_{0}$ of edge state. Here, we consider $N=11, \varphi=0.3\pi, d=3\lambda_0/4$.}\label{fig5}
\end{figure}

\textit{Dissipationless window}.---In Fig.~\ref{fig5}(a), we show the dissipation $\Gamma_{00}$ of a single edge state versus $d/\lambda_0$. Even though the chiral symmetry is broken for emitter spacings around $3\lambda_0/4$, the edge state can be dissipationless and is insensitive to emitter spacing. To confirm this conjecture, we rewrite the whole system as a non-Hermitian effective Hamiltonian in the diagonalized form $H_{\mathrm{eff}}=\sum_j (\tilde{E}_j - i \tilde{\Gamma}_j) |\tilde{\Psi}_j^R\rangle \langle \tilde{\Psi}_j^L|$ with the biorthogonal basis $ \langle \tilde{\Psi}_j^L|\tilde{\Psi}_{j'}^R\rangle =\delta_{jj'}$. A dissipationless window is found for $d=3\lambda_0/4$ with strong system-environment coupling, as shown in Fig.~\ref{fig5}(b). This window makes the dissipationless edge state robust to disorder in emitter positions~\cite{SupplementalMaterial}. The edge state has finite decay rate around $d=\lambda_0/4$. For emitter spacings $n \lambda_0/2$ ($n=0,1,2,\cdots$) where $H_{\mathrm{ph}}$ is zero, the edge state is more dissipative and shows higher sensitivity to disorder than $d=3\lambda_0/4$~\cite{SupplementalMaterial}.

\textit{Conclusions}.---System-environment interplay is fundamental for dissipative topological matter. In this work, we show that a 1D topological emitter array globally coupled to an electromagnetic environment exhibits interesting dissipative properties as the system-environment coupling varies. The energy spectrum width of the emitter array sets a critical value for the system-environment coupling and produces the dissipative topological phase transition (DTPT). The environment-modified topological edge states are stable and robust due to a dissipationless window in the emitter spacing. Our work paves an avenue for  electromagnetic control of topological matter with vacuum fields.

\begin{acknowledgments}We thank Christian Leefmans, Guo-Zhu Song, Clemens Gneiting, Yanming Che for critical readings and insightful comments. Y.X.L. is supported by the National Basic Research Program (973) of China under Grant No. 2017YFA0304304, the Key-Area Research and Development Program of GuangDong Province under Grant No. 2018B030326001, and NSFC under Grant No. 11874037. F.N. is supported in part by: Nippon Telegraph and Telephone Corporation (NTT) Research, Japan Science and Technology Agency (JST) (via the Quantum Leap Flagship Program (Q-LEAP), Moonshot R\&D Grant No. JPMJMS2061, and the Centers of Research Excellence in Science and Technology (CREST) Grant No. JPMJCR1676), Japan Society for the Promotion of Science (JSPS) (via the Grants-in-Aid for Scientific Research (KAKENHI) Grant No. JP20H00134, and the JSPS-RFBR Grant No. JPJSBP120194828), Army Research Office (ARO) (Grant No. W911NF-18-1-0358), the Asian Office of Aerospace Research and Development (AOARD) (via Grant No. FA2386-20-1-4069), and the Foundational Questions Institute (FQXi) (via Grant No. FQXi-IAF19-06).
\end{acknowledgments}


\begin{thebibliography}{83}%
\makeatletter
\providecommand \@ifxundefined [1]{%
 \@ifx{#1\undefined}
}%
\providecommand \@ifnum [1]{%
 \ifnum #1\expandafter \@firstoftwo
 \else \expandafter \@secondoftwo
 \fi
}%
\providecommand \@ifx [1]{%
 \ifx #1\expandafter \@firstoftwo
 \else \expandafter \@secondoftwo
 \fi
}%
\providecommand \natexlab [1]{#1}%
\providecommand \enquote  [1]{``#1''}%
\providecommand \bibnamefont  [1]{#1}%
\providecommand \bibfnamefont [1]{#1}%
\providecommand \citenamefont [1]{#1}%
\providecommand \href@noop [0]{\@secondoftwo}%
\providecommand \href [0]{\begingroup \@sanitize@url \@href}%
\providecommand \@href[1]{\@@startlink{#1}\@@href}%
\providecommand \@@href[1]{\endgroup#1\@@endlink}%
\providecommand \@sanitize@url [0]{\catcode `\\12\catcode `\$12\catcode
  `\&12\catcode `\#12\catcode `\^12\catcode `\_12\catcode `\%12\relax}%
\providecommand \@@startlink[1]{}%
\providecommand \@@endlink[0]{}%
\providecommand \url  [0]{\begingroup\@sanitize@url \@url }%
\providecommand \@url [1]{\endgroup\@href {#1}{\urlprefix }}%
\providecommand \urlprefix  [0]{URL }%
\providecommand \Eprint [0]{\href }%
\providecommand \doibase [0]{http://dx.doi.org/}%
\providecommand \selectlanguage [0]{\@gobble}%
\providecommand \bibinfo  [0]{\@secondoftwo}%
\providecommand \bibfield  [0]{\@secondoftwo}%
\providecommand \translation [1]{[#1]}%
\providecommand \BibitemOpen [0]{}%
\providecommand \bibitemStop [0]{}%
\providecommand \bibitemNoStop [0]{.\EOS\space}%
\providecommand \EOS [0]{\spacefactor3000\relax}%
\providecommand \BibitemShut  [1]{\csname bibitem#1\endcsname}%
\let\auto@bib@innerbib\@empty
\bibitem [{\citenamefont {Landig}\ \emph {et~al.}(2016)\citenamefont {Landig},
  \citenamefont {Hruby}, \citenamefont {Dogra}, \citenamefont {Landini},
  \citenamefont {Mottl}, \citenamefont {Donner},\ and\ \citenamefont
  {Esslinger}}]{landig2016quantum}%
  \BibitemOpen
  \bibfield  {author} {\bibinfo {author} {\bibfnamefont {R.}~\bibnamefont
  {Landig}}, \bibinfo {author} {\bibfnamefont {L.}~\bibnamefont {Hruby}},
  \bibinfo {author} {\bibfnamefont {N.}~\bibnamefont {Dogra}}, \bibinfo
  {author} {\bibfnamefont {M.}~\bibnamefont {Landini}}, \bibinfo {author}
  {\bibfnamefont {R.}~\bibnamefont {Mottl}}, \bibinfo {author} {\bibfnamefont
  {T.}~\bibnamefont {Donner}}, \ and\ \bibinfo {author} {\bibfnamefont
  {T.}~\bibnamefont {Esslinger}},\ }\bibfield  {title} {\emph {\bibinfo {title}
  {{Quantum phases from competing short- and long-range interactions in an
  optical lattice}},\ }}\href {\doibase 10.1038/nature17409} {\bibfield
  {journal} {\bibinfo  {journal} {Nature}\ }\textbf {\bibinfo {volume} {532}},\
  \bibinfo {pages} {476} (\bibinfo {year} {2016})}\BibitemShut {NoStop}%
\bibitem [{\citenamefont {Ashida}\ \emph {et~al.}(2020)\citenamefont {Ashida},
  \citenamefont {\ifmmode \dot{I}\else \.{I}\fi{}mamo\ifmmode~\breve{g}\else
  \u{g}\fi{}lu}, \citenamefont {Faist}, \citenamefont {Jaksch}, \citenamefont
  {Cavalleri},\ and\ \citenamefont {Demler}}]{PhysRevX.10.041027}%
  \BibitemOpen
  \bibfield  {author} {\bibinfo {author} {\bibfnamefont {Y.}~\bibnamefont
  {Ashida}}, \bibinfo {author} {\bibfnamefont {A.}~\bibnamefont {\ifmmode
  \dot{I}\else \.{I}\fi{}mamo\ifmmode~\breve{g}\else \u{g}\fi{}lu}}, \bibinfo
  {author} {\bibfnamefont {J.}~\bibnamefont {Faist}}, \bibinfo {author}
  {\bibfnamefont {D.}~\bibnamefont {Jaksch}}, \bibinfo {author} {\bibfnamefont
  {A.}~\bibnamefont {Cavalleri}}, \ and\ \bibinfo {author} {\bibfnamefont
  {E.}~\bibnamefont {Demler}},\ }\bibfield  {title} {\emph {\bibinfo {title}
  {{Quantum Electrodynamic Control of Matter: Cavity-Enhanced Ferroelectric
  Phase Transition}},\ }}\href {\doibase 10.1103/PhysRevX.10.041027} {\bibfield
   {journal} {\bibinfo  {journal} {Phys. Rev. X}\ }\textbf {\bibinfo {volume}
  {10}},\ \bibinfo {pages} {041027} (\bibinfo {year} {2020})}\BibitemShut
  {NoStop}%
\bibitem [{\citenamefont {Haroche}\ and\ \citenamefont
  {Raimond}(2006)}]{Haroche2006book}%
  \BibitemOpen
  \bibfield  {author} {\bibinfo {author} {\bibfnamefont {S.}~\bibnamefont
  {Haroche}}\ and\ \bibinfo {author} {\bibfnamefont {J.-M.}\ \bibnamefont
  {Raimond}},\ }\href@noop {} {\emph {\bibinfo {title} {Exploring the Quantum:
  Atoms, Cavities, and Photons}}}\ (\bibinfo  {publisher} {Oxford University
  Press},\ \bibinfo {year} {2006})\BibitemShut {NoStop}%
\bibitem [{\citenamefont {Gu}\ \emph {et~al.}(2017)\citenamefont {Gu},
  \citenamefont {Kockum}, \citenamefont {Miranowicz}, \citenamefont {Liu},\
  and\ \citenamefont {Nori}}]{gu2017microwave}%
  \BibitemOpen
  \bibfield  {author} {\bibinfo {author} {\bibfnamefont {X.}~\bibnamefont
  {Gu}}, \bibinfo {author} {\bibfnamefont {A.~F.}\ \bibnamefont {Kockum}},
  \bibinfo {author} {\bibfnamefont {A.}~\bibnamefont {Miranowicz}}, \bibinfo
  {author} {\bibfnamefont {Y.-X.}\ \bibnamefont {Liu}}, \ and\ \bibinfo
  {author} {\bibfnamefont {F.}~\bibnamefont {Nori}},\ }\bibfield  {title}
  {\emph {\bibinfo {title} {{Microwave photonics with superconducting quantum
  circuits}},\ }}\href {\doibase 10.1016/j.physrep.2017.10.002} {\bibfield
  {journal} {\bibinfo  {journal} {Phys. Rep.}\ }\textbf {\bibinfo {volume}
  {718-719}},\ \bibinfo {pages} {1} (\bibinfo {year} {2017})}\BibitemShut
  {NoStop}%
\bibitem [{\citenamefont {Kockum}\ \emph {et~al.}(2019)\citenamefont {Kockum},
  \citenamefont {Miranowicz}, \citenamefont {De~Liberato}, \citenamefont
  {Savasta},\ and\ \citenamefont {Nori}}]{kockum2019ultrastrong}%
  \BibitemOpen
  \bibfield  {author} {\bibinfo {author} {\bibfnamefont {A.~F.}\ \bibnamefont
  {Kockum}}, \bibinfo {author} {\bibfnamefont {A.}~\bibnamefont {Miranowicz}},
  \bibinfo {author} {\bibfnamefont {S.}~\bibnamefont {De~Liberato}}, \bibinfo
  {author} {\bibfnamefont {S.}~\bibnamefont {Savasta}}, \ and\ \bibinfo
  {author} {\bibfnamefont {F.}~\bibnamefont {Nori}},\ }\bibfield  {title}
  {\emph {\bibinfo {title} {{Ultrastrong coupling between light and matter}},\
  }}\href {\doibase 10.1038/s42254-018-0006-2} {\bibfield  {journal} {\bibinfo
  {journal} {Nat. Rev. Phys.}\ }\textbf {\bibinfo {volume} {1}},\ \bibinfo
  {pages} {19} (\bibinfo {year} {2019})}\BibitemShut {NoStop}%
\bibitem [{\citenamefont {Mivehvar}\ \emph {et~al.}(2021)\citenamefont
  {Mivehvar}, \citenamefont {Piazza}, \citenamefont {Donner},\ and\
  \citenamefont {Ritsch}}]{mivehvar2021cavity}%
  \BibitemOpen
  \bibfield  {author} {\bibinfo {author} {\bibfnamefont {F.}~\bibnamefont
  {Mivehvar}}, \bibinfo {author} {\bibfnamefont {F.}~\bibnamefont {Piazza}},
  \bibinfo {author} {\bibfnamefont {T.}~\bibnamefont {Donner}}, \ and\ \bibinfo
  {author} {\bibfnamefont {H.}~\bibnamefont {Ritsch}},\ }\bibfield  {title}
  {\emph {\bibinfo {title} {{Cavity QED with Quantum Gases: New Paradigms in
  Many-Body Physics}},\ }}\href {https://arxiv.org/abs/2102.04473} {\bibfield
  {journal} {\bibinfo  {journal} {arXiv:2102.04473}\ } (\bibinfo {year}
  {2021})}\BibitemShut {NoStop}%
\bibitem [{\citenamefont {Schlawin}\ \emph {et~al.}(2019)\citenamefont
  {Schlawin}, \citenamefont {Cavalleri},\ and\ \citenamefont
  {Jaksch}}]{PhysRevLett.122.133602}%
  \BibitemOpen
  \bibfield  {author} {\bibinfo {author} {\bibfnamefont {F.}~\bibnamefont
  {Schlawin}}, \bibinfo {author} {\bibfnamefont {A.}~\bibnamefont {Cavalleri}},
  \ and\ \bibinfo {author} {\bibfnamefont {D.}~\bibnamefont {Jaksch}},\
  }\bibfield  {title} {\emph {\bibinfo {title} {{Cavity-Mediated
  Electron-Photon Superconductivity}},\ }}\href {\doibase
  10.1103/PhysRevLett.122.133602} {\bibfield  {journal} {\bibinfo  {journal}
  {Phys. Rev. Lett.}\ }\textbf {\bibinfo {volume} {122}},\ \bibinfo {pages}
  {133602} (\bibinfo {year} {2019})}\BibitemShut {NoStop}%
\bibitem [{\citenamefont {Curtis}\ \emph {et~al.}(2019)\citenamefont {Curtis},
  \citenamefont {Raines}, \citenamefont {Allocca}, \citenamefont {Hafezi},\
  and\ \citenamefont {Galitski}}]{PhysRevLett.122.167002}%
  \BibitemOpen
  \bibfield  {author} {\bibinfo {author} {\bibfnamefont {J.~B.}\ \bibnamefont
  {Curtis}}, \bibinfo {author} {\bibfnamefont {Z.~M.}\ \bibnamefont {Raines}},
  \bibinfo {author} {\bibfnamefont {A.~A.}\ \bibnamefont {Allocca}}, \bibinfo
  {author} {\bibfnamefont {M.}~\bibnamefont {Hafezi}}, \ and\ \bibinfo {author}
  {\bibfnamefont {V.~M.}\ \bibnamefont {Galitski}},\ }\bibfield  {title} {\emph
  {\bibinfo {title} {{Cavity Quantum Eliashberg Enhancement of
  Superconductivity}},\ }}\href {\doibase 10.1103/PhysRevLett.122.167002}
  {\bibfield  {journal} {\bibinfo  {journal} {Phys. Rev. Lett.}\ }\textbf
  {\bibinfo {volume} {122}},\ \bibinfo {pages} {167002} (\bibinfo {year}
  {2019})}\BibitemShut {NoStop}%
\bibitem [{\citenamefont {Thomas}\ \emph {et~al.}(2019)\citenamefont {Thomas},
  \citenamefont {Devaux}, \citenamefont {Nagarajan}, \citenamefont {Chervy},
  \citenamefont {Seidel}, \citenamefont {Hagenm{\"u}ller}, \citenamefont
  {Sch{\"u}tz}, \citenamefont {Schachenmayer}, \citenamefont {Genet},
  \citenamefont {Pupillo},\ and\ \citenamefont
  {Ebbesen}}]{thomas2019exploring}%
  \BibitemOpen
  \bibfield  {author} {\bibinfo {author} {\bibfnamefont {A.}~\bibnamefont
  {Thomas}}, \bibinfo {author} {\bibfnamefont {E.}~\bibnamefont {Devaux}},
  \bibinfo {author} {\bibfnamefont {K.}~\bibnamefont {Nagarajan}}, \bibinfo
  {author} {\bibfnamefont {T.}~\bibnamefont {Chervy}}, \bibinfo {author}
  {\bibfnamefont {M.}~\bibnamefont {Seidel}}, \bibinfo {author} {\bibfnamefont
  {D.}~\bibnamefont {Hagenm{\"u}ller}}, \bibinfo {author} {\bibfnamefont
  {S.}~\bibnamefont {Sch{\"u}tz}}, \bibinfo {author} {\bibfnamefont
  {J.}~\bibnamefont {Schachenmayer}}, \bibinfo {author} {\bibfnamefont
  {C.}~\bibnamefont {Genet}}, \bibinfo {author} {\bibfnamefont
  {G.}~\bibnamefont {Pupillo}}, \ and\ \bibinfo {author} {\bibfnamefont
  {T.~W.}\ \bibnamefont {Ebbesen}},\ }\bibfield  {title} {\emph {\bibinfo
  {title} {{Exploring Superconductivity under Strong Coupling with the Vacuum
  Electromagnetic Field}},\ }}\href {https://arxiv.org/abs/1911.01459}
  {\bibfield  {journal} {\bibinfo  {journal} {arXiv:1911.01459}\ } (\bibinfo
  {year} {2019})}\BibitemShut {NoStop}%
\bibitem [{\citenamefont {Garcia-Vidal}\ \emph {et~al.}(2021)\citenamefont
  {Garcia-Vidal}, \citenamefont {Ciuti},\ and\ \citenamefont
  {Ebbesen}}]{Garcia2021}%
  \BibitemOpen
  \bibfield  {author} {\bibinfo {author} {\bibfnamefont {F.~J.}\ \bibnamefont
  {Garcia-Vidal}}, \bibinfo {author} {\bibfnamefont {C.}~\bibnamefont {Ciuti}},
  \ and\ \bibinfo {author} {\bibfnamefont {T.~W.}\ \bibnamefont {Ebbesen}},\
  }\bibfield  {title} {\emph {\bibinfo {title} {{Manipulating matter by strong
  coupling to vacuum fields}},\ }}\href {\doibase 10.1126/science.abd0336}
  {\bibfield  {journal} {\bibinfo  {journal} {Science}\ }\textbf {\bibinfo
  {volume} {373}},\ \bibinfo {pages} {eabd0336} (\bibinfo {year}
  {2021})}\BibitemShut {NoStop}%
\bibitem [{\citenamefont {Gonz{\'a}lez-Tudela}\ \emph
  {et~al.}(2015)\citenamefont {Gonz{\'a}lez-Tudela}, \citenamefont {Hung},
  \citenamefont {Chang}, \citenamefont {Cirac},\ and\ \citenamefont
  {Kimble}}]{gonzalez2015subwavelength}%
  \BibitemOpen
  \bibfield  {author} {\bibinfo {author} {\bibfnamefont {A.}~\bibnamefont
  {Gonz{\'a}lez-Tudela}}, \bibinfo {author} {\bibfnamefont {C.-L.}\
  \bibnamefont {Hung}}, \bibinfo {author} {\bibfnamefont {D.~E.}\ \bibnamefont
  {Chang}}, \bibinfo {author} {\bibfnamefont {J.~I.}\ \bibnamefont {Cirac}}, \
  and\ \bibinfo {author} {\bibfnamefont {H.}~\bibnamefont {Kimble}},\
  }\bibfield  {title} {\emph {\bibinfo {title} {Subwavelength vacuum lattices
  and atom--atom interactions in two-dimensional photonic crystals},\ }}\href
  {\doibase 10.1038/nphoton.2015.54} {\bibfield  {journal} {\bibinfo  {journal}
  {Nat. Photonics}\ }\textbf {\bibinfo {volume} {9}},\ \bibinfo {pages} {320}
  (\bibinfo {year} {2015})}\BibitemShut {NoStop}%
\bibitem [{\citenamefont {Perczel}\ \emph {et~al.}(2020)\citenamefont
  {Perczel}, \citenamefont {Borregaard}, \citenamefont {Chang}, \citenamefont
  {Yelin},\ and\ \citenamefont {Lukin}}]{PhysRevLett.124.083603}%
  \BibitemOpen
  \bibfield  {author} {\bibinfo {author} {\bibfnamefont {J.}~\bibnamefont
  {Perczel}}, \bibinfo {author} {\bibfnamefont {J.}~\bibnamefont {Borregaard}},
  \bibinfo {author} {\bibfnamefont {D.~E.}\ \bibnamefont {Chang}}, \bibinfo
  {author} {\bibfnamefont {S.~F.}\ \bibnamefont {Yelin}}, \ and\ \bibinfo
  {author} {\bibfnamefont {M.~D.}\ \bibnamefont {Lukin}},\ }\bibfield  {title}
  {\emph {\bibinfo {title} {{Topological Quantum Optics Using Atomlike Emitter
  Arrays Coupled to Photonic Crystals}},\ }}\href {\doibase
  10.1103/PhysRevLett.124.083603} {\bibfield  {journal} {\bibinfo  {journal}
  {Phys. Rev. Lett.}\ }\textbf {\bibinfo {volume} {124}},\ \bibinfo {pages}
  {083603} (\bibinfo {year} {2020})}\BibitemShut {NoStop}%
\bibitem [{\citenamefont {Rui}\ \emph {et~al.}(2020)\citenamefont {Rui},
  \citenamefont {Wei}, \citenamefont {Rubio-Abadal}, \citenamefont {Hollerith},
  \citenamefont {Zeiher}, \citenamefont {Stamper-Kurn}, \citenamefont {Gross},\
  and\ \citenamefont {Bloch}}]{rui2020subradiant}%
  \BibitemOpen
  \bibfield  {author} {\bibinfo {author} {\bibfnamefont {J.}~\bibnamefont
  {Rui}}, \bibinfo {author} {\bibfnamefont {D.}~\bibnamefont {Wei}}, \bibinfo
  {author} {\bibfnamefont {A.}~\bibnamefont {Rubio-Abadal}}, \bibinfo {author}
  {\bibfnamefont {S.}~\bibnamefont {Hollerith}}, \bibinfo {author}
  {\bibfnamefont {J.}~\bibnamefont {Zeiher}}, \bibinfo {author} {\bibfnamefont
  {D.~M.}\ \bibnamefont {Stamper-Kurn}}, \bibinfo {author} {\bibfnamefont
  {C.}~\bibnamefont {Gross}}, \ and\ \bibinfo {author} {\bibfnamefont
  {I.}~\bibnamefont {Bloch}},\ }\bibfield  {title} {\emph {\bibinfo {title} {A
  subradiant optical mirror formed by a single structured atomic layer},\
  }}\href {\doibase 10.1038/s41586-020-2463-x} {\bibfield  {journal} {\bibinfo
  {journal} {Nature}\ }\textbf {\bibinfo {volume} {583}},\ \bibinfo {pages}
  {369} (\bibinfo {year} {2020})}\BibitemShut {NoStop}%
\bibitem [{\citenamefont {Trif}\ and\ \citenamefont
  {Simon}(2019)}]{PhysRevLett.122.236803}%
  \BibitemOpen
  \bibfield  {author} {\bibinfo {author} {\bibfnamefont {M.}~\bibnamefont
  {Trif}}\ and\ \bibinfo {author} {\bibfnamefont {P.}~\bibnamefont {Simon}},\
  }\bibfield  {title} {\emph {\bibinfo {title} {{Braiding of Majorana Fermions
  in a Cavity}},\ }}\href {\doibase 10.1103/PhysRevLett.122.236803} {\bibfield
  {journal} {\bibinfo  {journal} {Phys. Rev. Lett.}\ }\textbf {\bibinfo
  {volume} {122}},\ \bibinfo {pages} {236803} (\bibinfo {year}
  {2019})}\BibitemShut {NoStop}%
\bibitem [{\citenamefont {Nie}\ and\ \citenamefont
  {Liu}(2020)}]{PhysRevResearch.2.012076}%
  \BibitemOpen
  \bibfield  {author} {\bibinfo {author} {\bibfnamefont {W.}~\bibnamefont
  {Nie}}\ and\ \bibinfo {author} {\bibfnamefont {Y.-X.}\ \bibnamefont {Liu}},\
  }\bibfield  {title} {\emph {\bibinfo {title} {Bandgap-assisted quantum
  control of topological edge states in a cavity},\ }}\href {\doibase
  10.1103/PhysRevResearch.2.012076} {\bibfield  {journal} {\bibinfo  {journal}
  {Phys. Rev. Research}\ }\textbf {\bibinfo {volume} {2}},\ \bibinfo {pages}
  {012076(R)} (\bibinfo {year} {2020})}\BibitemShut {NoStop}%
\bibitem [{\citenamefont {Mann}\ and\ \citenamefont
  {Mariani}(2020)}]{mann2020topological}%
  \BibitemOpen
  \bibfield  {author} {\bibinfo {author} {\bibfnamefont {C.-R.}\ \bibnamefont
  {Mann}}\ and\ \bibinfo {author} {\bibfnamefont {E.}~\bibnamefont {Mariani}},\
  }\bibfield  {title} {\emph {\bibinfo {title} {{Topological transitions
  induced by cavity-mediated interactions in photonic valley-Hall
  metasurfaces}},\ }}\href {https://arxiv.org/abs/2010.01636} {\bibfield
  {journal} {\bibinfo  {journal} {arXiv:2010.01636}\ } (\bibinfo {year}
  {2020})}\BibitemShut {NoStop}%
\bibitem [{\citenamefont {Mann}\ \emph {et~al.}(2020)\citenamefont {Mann},
  \citenamefont {Horsley},\ and\ \citenamefont {Mariani}}]{mann2020tunable}%
  \BibitemOpen
  \bibfield  {author} {\bibinfo {author} {\bibfnamefont {C.-R.}\ \bibnamefont
  {Mann}}, \bibinfo {author} {\bibfnamefont {S.~A.}\ \bibnamefont {Horsley}}, \
  and\ \bibinfo {author} {\bibfnamefont {E.}~\bibnamefont {Mariani}},\
  }\bibfield  {title} {\emph {\bibinfo {title} {Tunable pseudo-magnetic fields
  for polaritons in strained metasurfaces},\ }}\href {\doibase
  10.1038/s41566-020-0688-8} {\bibfield  {journal} {\bibinfo  {journal} {Nat.
  Photonics}\ }\textbf {\bibinfo {volume} {14}},\ \bibinfo {pages} {669}
  (\bibinfo {year} {2020})}\BibitemShut {NoStop}%
\bibitem [{\citenamefont {Diehl}\ \emph {et~al.}(2011)\citenamefont {Diehl},
  \citenamefont {Rico}, \citenamefont {Baranov},\ and\ \citenamefont
  {Zoller}}]{diehl2011topology}%
  \BibitemOpen
  \bibfield  {author} {\bibinfo {author} {\bibfnamefont {S.}~\bibnamefont
  {Diehl}}, \bibinfo {author} {\bibfnamefont {E.}~\bibnamefont {Rico}},
  \bibinfo {author} {\bibfnamefont {M.~A.}\ \bibnamefont {Baranov}}, \ and\
  \bibinfo {author} {\bibfnamefont {P.}~\bibnamefont {Zoller}},\ }\bibfield
  {title} {\emph {\bibinfo {title} {{Topology by dissipation in atomic quantum
  wires}},\ }}\href {\doibase 10.1038/nphys2106} {\bibfield  {journal}
  {\bibinfo  {journal} {Nat. Phys.}\ }\textbf {\bibinfo {volume} {7}},\
  \bibinfo {pages} {971} (\bibinfo {year} {2011})}\BibitemShut {NoStop}%
\bibitem [{\citenamefont {Goldstein}\ and\ \citenamefont
  {Chamon}(2011)}]{PhysRevB.84.205109}%
  \BibitemOpen
  \bibfield  {author} {\bibinfo {author} {\bibfnamefont {G.}~\bibnamefont
  {Goldstein}}\ and\ \bibinfo {author} {\bibfnamefont {C.}~\bibnamefont
  {Chamon}},\ }\bibfield  {title} {\emph {\bibinfo {title} {{Decay rates for
  topological memories encoded with Majorana fermions}},\ }}\href {\doibase
  10.1103/PhysRevB.84.205109} {\bibfield  {journal} {\bibinfo  {journal} {Phys.
  Rev. B}\ }\textbf {\bibinfo {volume} {84}},\ \bibinfo {pages} {205109}
  (\bibinfo {year} {2011})}\BibitemShut {NoStop}%
\bibitem [{\citenamefont {Budich}\ \emph {et~al.}(2012)\citenamefont {Budich},
  \citenamefont {Walter},\ and\ \citenamefont
  {Trauzettel}}]{PhysRevB.85.121405}%
  \BibitemOpen
  \bibfield  {author} {\bibinfo {author} {\bibfnamefont {J.~C.}\ \bibnamefont
  {Budich}}, \bibinfo {author} {\bibfnamefont {S.}~\bibnamefont {Walter}}, \
  and\ \bibinfo {author} {\bibfnamefont {B.}~\bibnamefont {Trauzettel}},\
  }\bibfield  {title} {\emph {\bibinfo {title} {{Failure of protection of
  Majorana based qubits against decoherence}},\ }}\href {\doibase
  10.1103/PhysRevB.85.121405} {\bibfield  {journal} {\bibinfo  {journal} {Phys.
  Rev. B}\ }\textbf {\bibinfo {volume} {85}},\ \bibinfo {pages} {121405}
  (\bibinfo {year} {2012})}\BibitemShut {NoStop}%
\bibitem [{\citenamefont {Mazza}\ \emph {et~al.}(2013)\citenamefont {Mazza},
  \citenamefont {Rizzi}, \citenamefont {Lukin},\ and\ \citenamefont
  {Cirac}}]{PhysRevB.88.205142}%
  \BibitemOpen
  \bibfield  {author} {\bibinfo {author} {\bibfnamefont {L.}~\bibnamefont
  {Mazza}}, \bibinfo {author} {\bibfnamefont {M.}~\bibnamefont {Rizzi}},
  \bibinfo {author} {\bibfnamefont {M.~D.}\ \bibnamefont {Lukin}}, \ and\
  \bibinfo {author} {\bibfnamefont {J.~I.}\ \bibnamefont {Cirac}},\ }\bibfield
  {title} {\emph {\bibinfo {title} {{Robustness of quantum memories based on
  Majorana zero modes}},\ }}\href {\doibase 10.1103/PhysRevB.88.205142}
  {\bibfield  {journal} {\bibinfo  {journal} {Phys. Rev. B}\ }\textbf {\bibinfo
  {volume} {88}},\ \bibinfo {pages} {205142} (\bibinfo {year}
  {2013})}\BibitemShut {NoStop}%
\bibitem [{\citenamefont {Bardyn}\ \emph {et~al.}(2013)\citenamefont {Bardyn},
  \citenamefont {Baranov}, \citenamefont {Kraus}, \citenamefont {Rico},
  \citenamefont {{\.I}mamo{\u{g}}lu}, \citenamefont {Zoller},\ and\
  \citenamefont {Diehl}}]{bardyn2013topology}%
  \BibitemOpen
  \bibfield  {author} {\bibinfo {author} {\bibfnamefont {C.-E.}\ \bibnamefont
  {Bardyn}}, \bibinfo {author} {\bibfnamefont {M.~A.}\ \bibnamefont {Baranov}},
  \bibinfo {author} {\bibfnamefont {C.~V.}\ \bibnamefont {Kraus}}, \bibinfo
  {author} {\bibfnamefont {E.}~\bibnamefont {Rico}}, \bibinfo {author}
  {\bibfnamefont {A.}~\bibnamefont {{\.I}mamo{\u{g}}lu}}, \bibinfo {author}
  {\bibfnamefont {P.}~\bibnamefont {Zoller}}, \ and\ \bibinfo {author}
  {\bibfnamefont {S.}~\bibnamefont {Diehl}},\ }\bibfield  {title} {\emph
  {\bibinfo {title} {{Topology by dissipation}},\ }}\href {\doibase
  10.1088/1367-2630/15/8/085001} {\bibfield  {journal} {\bibinfo  {journal}
  {New J. Phys.}\ }\textbf {\bibinfo {volume} {15}},\ \bibinfo {pages} {085001}
  (\bibinfo {year} {2013})}\BibitemShut {NoStop}%
\bibitem [{\citenamefont {Shen}\ \emph {et~al.}(2014)\citenamefont {Shen},
  \citenamefont {Wang},\ and\ \citenamefont {Yi}}]{shen2014hall}%
  \BibitemOpen
  \bibfield  {author} {\bibinfo {author} {\bibfnamefont {H.~Z.}\ \bibnamefont
  {Shen}}, \bibinfo {author} {\bibfnamefont {W.}~\bibnamefont {Wang}}, \ and\
  \bibinfo {author} {\bibfnamefont {X.~X.}\ \bibnamefont {Yi}},\ }\bibfield
  {title} {\emph {\bibinfo {title} {{Hall conductance and topological invariant
  for open systems}},\ }}\href {\doibase 10.1038/srep06455} {\bibfield
  {journal} {\bibinfo  {journal} {Sci. Rep.}\ }\textbf {\bibinfo {volume}
  {4}},\ \bibinfo {pages} {1} (\bibinfo {year} {2014})}\BibitemShut {NoStop}%
\bibitem [{\citenamefont {Hu}\ \emph {et~al.}(2015)\citenamefont {Hu},
  \citenamefont {Cai}, \citenamefont {Baranov},\ and\ \citenamefont
  {Zoller}}]{PhysRevB.92.165118}%
  \BibitemOpen
  \bibfield  {author} {\bibinfo {author} {\bibfnamefont {Y.}~\bibnamefont
  {Hu}}, \bibinfo {author} {\bibfnamefont {Z.}~\bibnamefont {Cai}}, \bibinfo
  {author} {\bibfnamefont {M.~A.}\ \bibnamefont {Baranov}}, \ and\ \bibinfo
  {author} {\bibfnamefont {P.}~\bibnamefont {Zoller}},\ }\bibfield  {title}
  {\emph {\bibinfo {title} {{Majorana fermions in noisy Kitaev wires}},\
  }}\href {\doibase 10.1103/PhysRevB.92.165118} {\bibfield  {journal} {\bibinfo
   {journal} {Phys. Rev. B}\ }\textbf {\bibinfo {volume} {92}},\ \bibinfo
  {pages} {165118} (\bibinfo {year} {2015})}\BibitemShut {NoStop}%
\bibitem [{\citenamefont {Linzner}\ \emph {et~al.}(2016)\citenamefont
  {Linzner}, \citenamefont {Wawer}, \citenamefont {Grusdt},\ and\ \citenamefont
  {Fleischhauer}}]{PhysRevB.94.201105}%
  \BibitemOpen
  \bibfield  {author} {\bibinfo {author} {\bibfnamefont {D.}~\bibnamefont
  {Linzner}}, \bibinfo {author} {\bibfnamefont {L.}~\bibnamefont {Wawer}},
  \bibinfo {author} {\bibfnamefont {F.}~\bibnamefont {Grusdt}}, \ and\ \bibinfo
  {author} {\bibfnamefont {M.}~\bibnamefont {Fleischhauer}},\ }\bibfield
  {title} {\emph {\bibinfo {title} {{Reservoir-induced Thouless pumping and
  symmetry-protected topological order in open quantum chains}},\ }}\href
  {\doibase 10.1103/PhysRevB.94.201105} {\bibfield  {journal} {\bibinfo
  {journal} {Phys. Rev. B}\ }\textbf {\bibinfo {volume} {94}},\ \bibinfo
  {pages} {201105} (\bibinfo {year} {2016})}\BibitemShut {NoStop}%
\bibitem [{\citenamefont {Else}\ \emph {et~al.}(2017)\citenamefont {Else},
  \citenamefont {Fendley}, \citenamefont {Kemp},\ and\ \citenamefont
  {Nayak}}]{PhysRevX.7.041062}%
  \BibitemOpen
  \bibfield  {author} {\bibinfo {author} {\bibfnamefont {D.~V.}\ \bibnamefont
  {Else}}, \bibinfo {author} {\bibfnamefont {P.}~\bibnamefont {Fendley}},
  \bibinfo {author} {\bibfnamefont {J.}~\bibnamefont {Kemp}}, \ and\ \bibinfo
  {author} {\bibfnamefont {C.}~\bibnamefont {Nayak}},\ }\bibfield  {title}
  {\emph {\bibinfo {title} {{Prethermal Strong Zero Modes and Topological
  Qubits}},\ }}\href {\doibase 10.1103/PhysRevX.7.041062} {\bibfield  {journal}
  {\bibinfo  {journal} {Phys. Rev. X}\ }\textbf {\bibinfo {volume} {7}},\
  \bibinfo {pages} {041062} (\bibinfo {year} {2017})}\BibitemShut {NoStop}%
\bibitem [{\citenamefont {Carollo}\ \emph {et~al.}(2018)\citenamefont
  {Carollo}, \citenamefont {Spagnolo},\ and\ \citenamefont
  {Valenti}}]{carollo2018uhlmann}%
  \BibitemOpen
  \bibfield  {author} {\bibinfo {author} {\bibfnamefont {A.}~\bibnamefont
  {Carollo}}, \bibinfo {author} {\bibfnamefont {B.}~\bibnamefont {Spagnolo}}, \
  and\ \bibinfo {author} {\bibfnamefont {D.}~\bibnamefont {Valenti}},\
  }\bibfield  {title} {\emph {\bibinfo {title} {{Uhlmann curvature in
  dissipative phase transitions}},\ }}\href {\doibase
  10.1038/s41598-018-27362-9} {\bibfield  {journal} {\bibinfo  {journal} {Sci.
  Rep.}\ }\textbf {\bibinfo {volume} {8}},\ \bibinfo {pages} {1} (\bibinfo
  {year} {2018})}\BibitemShut {NoStop}%
\bibitem [{\citenamefont {Kastoryano}\ and\ \citenamefont
  {Rudner}(2019)}]{PhysRevB.99.125118}%
  \BibitemOpen
  \bibfield  {author} {\bibinfo {author} {\bibfnamefont {M.~J.}\ \bibnamefont
  {Kastoryano}}\ and\ \bibinfo {author} {\bibfnamefont {M.~S.}\ \bibnamefont
  {Rudner}},\ }\bibfield  {title} {\emph {\bibinfo {title} {Topological
  transport in the steady state of a quantum particle with dissipation},\
  }}\href {\doibase 10.1103/PhysRevB.99.125118} {\bibfield  {journal} {\bibinfo
   {journal} {Phys. Rev. B}\ }\textbf {\bibinfo {volume} {99}},\ \bibinfo
  {pages} {125118} (\bibinfo {year} {2019})}\BibitemShut {NoStop}%
\bibitem [{\citenamefont {Weisbrich}\ \emph {et~al.}(2019)\citenamefont
  {Weisbrich}, \citenamefont {Belzig},\ and\ \citenamefont
  {Rastelli}}]{weisbrich2019decoherence}%
  \BibitemOpen
  \bibfield  {author} {\bibinfo {author} {\bibfnamefont {H.}~\bibnamefont
  {Weisbrich}}, \bibinfo {author} {\bibfnamefont {W.}~\bibnamefont {Belzig}}, \
  and\ \bibinfo {author} {\bibfnamefont {G.}~\bibnamefont {Rastelli}},\
  }\bibfield  {title} {\emph {\bibinfo {title} {{Decoherence and relaxation of
  topological states in extended quantum Ising models}},\ }}\href {\doibase
  10.21468/SciPostPhys.6.3.037} {\bibfield  {journal} {\bibinfo  {journal}
  {SciPost Phys.}\ }\textbf {\bibinfo {volume} {6}},\ \bibinfo {pages} {37}
  (\bibinfo {year} {2019})}\BibitemShut {NoStop}%
\bibitem [{\citenamefont {Carollo}\ \emph {et~al.}(2020)\citenamefont
  {Carollo}, \citenamefont {Valenti},\ and\ \citenamefont
  {Spagnolo}}]{carollo2020geometry}%
  \BibitemOpen
  \bibfield  {author} {\bibinfo {author} {\bibfnamefont {A.}~\bibnamefont
  {Carollo}}, \bibinfo {author} {\bibfnamefont {D.}~\bibnamefont {Valenti}}, \
  and\ \bibinfo {author} {\bibfnamefont {B.}~\bibnamefont {Spagnolo}},\
  }\bibfield  {title} {\emph {\bibinfo {title} {{Geometry of quantum phase
  transitions}},\ }}\href {\doibase 10.1016/j.physrep.2019.11.002} {\bibfield
  {journal} {\bibinfo  {journal} {Phys. Rep.}\ }\textbf {\bibinfo {volume}
  {838}},\ \bibinfo {pages} {1} (\bibinfo {year} {2020})}\BibitemShut {NoStop}%
\bibitem [{\citenamefont {Huang}\ \emph {et~al.}(2020)\citenamefont {Huang},
  \citenamefont {Yang},\ and\ \citenamefont {Zhang}}]{PhysRevB.102.165116}%
  \BibitemOpen
  \bibfield  {author} {\bibinfo {author} {\bibfnamefont {Y.-W.}\ \bibnamefont
  {Huang}}, \bibinfo {author} {\bibfnamefont {P.-Y.}\ \bibnamefont {Yang}}, \
  and\ \bibinfo {author} {\bibfnamefont {W.-M.}\ \bibnamefont {Zhang}},\
  }\bibfield  {title} {\emph {\bibinfo {title} {{Quantum theory of dissipative
  topological systems}},\ }}\href {\doibase 10.1103/PhysRevB.102.165116}
  {\bibfield  {journal} {\bibinfo  {journal} {Phys. Rev. B}\ }\textbf {\bibinfo
  {volume} {102}},\ \bibinfo {pages} {165116} (\bibinfo {year}
  {2020})}\BibitemShut {NoStop}%
\bibitem [{\citenamefont {Gneiting}\ \emph {et~al.}(2020)\citenamefont
  {Gneiting}, \citenamefont {Koottandavida}, \citenamefont {Rozhkov},\ and\
  \citenamefont {Nori}}]{gneiting2020unraveling}%
  \BibitemOpen
  \bibfield  {author} {\bibinfo {author} {\bibfnamefont {C.}~\bibnamefont
  {Gneiting}}, \bibinfo {author} {\bibfnamefont {A.}~\bibnamefont
  {Koottandavida}}, \bibinfo {author} {\bibfnamefont {A.~V.}\ \bibnamefont
  {Rozhkov}}, \ and\ \bibinfo {author} {\bibfnamefont {F.}~\bibnamefont
  {Nori}},\ }\bibfield  {title} {\emph {\bibinfo {title} {{Unraveling the
  topology of dissipative quantum systems}},\ }}\href
  {https://arxiv.org/abs/2007.05960} {\bibfield  {journal} {\bibinfo  {journal}
  {arXiv:2007.05960}\ } (\bibinfo {year} {2020})}\BibitemShut {NoStop}%
\bibitem [{\citenamefont {Leefmans}\ \emph {et~al.}(2021)\citenamefont
  {Leefmans}, \citenamefont {Dutt}, \citenamefont {Williams}, \citenamefont
  {Yuan}, \citenamefont {Parto}, \citenamefont {Nori}, \citenamefont {Fan},\
  and\ \citenamefont {Marandi}}]{Leefmans2021}%
  \BibitemOpen
  \bibfield  {author} {\bibinfo {author} {\bibfnamefont {C.}~\bibnamefont
  {Leefmans}}, \bibinfo {author} {\bibfnamefont {A.}~\bibnamefont {Dutt}},
  \bibinfo {author} {\bibfnamefont {J.}~\bibnamefont {Williams}}, \bibinfo
  {author} {\bibfnamefont {L.}~\bibnamefont {Yuan}}, \bibinfo {author}
  {\bibfnamefont {M.}~\bibnamefont {Parto}}, \bibinfo {author} {\bibfnamefont
  {F.}~\bibnamefont {Nori}}, \bibinfo {author} {\bibfnamefont {S.}~\bibnamefont
  {Fan}}, \ and\ \bibinfo {author} {\bibfnamefont {A.}~\bibnamefont
  {Marandi}},\ }\bibfield  {title} {\emph {\bibinfo {title} {{Photonic
  Topological Dissipation in a Time-Multiplexed Resonator Network}},\ }}\href
  {https://arxiv.org/abs/2104.05213} {\bibfield  {journal} {\bibinfo  {journal}
  {arXiv:2104.05213}\ } (\bibinfo {year} {2021})}\BibitemShut {NoStop}%
\bibitem [{\citenamefont {Atala}\ \emph {et~al.}(2013)\citenamefont {Atala},
  \citenamefont {Aidelsburger}, \citenamefont {Barreiro}, \citenamefont
  {Abanin}, \citenamefont {Kitagawa}, \citenamefont {Demler},\ and\
  \citenamefont {Bloch}}]{atala2013direct}%
  \BibitemOpen
  \bibfield  {author} {\bibinfo {author} {\bibfnamefont {M.}~\bibnamefont
  {Atala}}, \bibinfo {author} {\bibfnamefont {M.}~\bibnamefont {Aidelsburger}},
  \bibinfo {author} {\bibfnamefont {J.~T.}\ \bibnamefont {Barreiro}}, \bibinfo
  {author} {\bibfnamefont {D.}~\bibnamefont {Abanin}}, \bibinfo {author}
  {\bibfnamefont {T.}~\bibnamefont {Kitagawa}}, \bibinfo {author}
  {\bibfnamefont {E.}~\bibnamefont {Demler}}, \ and\ \bibinfo {author}
  {\bibfnamefont {I.}~\bibnamefont {Bloch}},\ }\bibfield  {title} {\emph
  {\bibinfo {title} {{Direct measurement of the Zak phase in topological Bloch
  bands}},\ }}\href {\doibase 10.1038/nphys2790} {\bibfield  {journal}
  {\bibinfo  {journal} {Nat. Phys.}\ }\textbf {\bibinfo {volume} {9}},\
  \bibinfo {pages} {795} (\bibinfo {year} {2013})}\BibitemShut {NoStop}%
\bibitem [{\citenamefont {Zhang}\ \emph {et~al.}(2018)\citenamefont {Zhang},
  \citenamefont {Zhu}, \citenamefont {Zhao}, \citenamefont {Yan},\ and\
  \citenamefont {Zhu}}]{zhang2019topological}%
  \BibitemOpen
  \bibfield  {author} {\bibinfo {author} {\bibfnamefont {D.-W.}\ \bibnamefont
  {Zhang}}, \bibinfo {author} {\bibfnamefont {Y.-Q.}\ \bibnamefont {Zhu}},
  \bibinfo {author} {\bibfnamefont {Y.~X.}\ \bibnamefont {Zhao}}, \bibinfo
  {author} {\bibfnamefont {H.}~\bibnamefont {Yan}}, \ and\ \bibinfo {author}
  {\bibfnamefont {S.-L.}\ \bibnamefont {Zhu}},\ }\bibfield  {title} {\emph
  {\bibinfo {title} {{Topological quantum matter with cold atoms}},\ }}\href
  {\doibase 10.1080/00018732.2019.1594094} {\bibfield  {journal} {\bibinfo
  {journal} {Adv. Phys.}\ }\textbf {\bibinfo {volume} {67}},\ \bibinfo {pages}
  {253} (\bibinfo {year} {2018})}\BibitemShut {NoStop}%
\bibitem [{\citenamefont {Cooper}\ \emph {et~al.}(2019)\citenamefont {Cooper},
  \citenamefont {Dalibard},\ and\ \citenamefont
  {Spielman}}]{RevModPhys.91.015005}%
  \BibitemOpen
  \bibfield  {author} {\bibinfo {author} {\bibfnamefont {N.~R.}\ \bibnamefont
  {Cooper}}, \bibinfo {author} {\bibfnamefont {J.}~\bibnamefont {Dalibard}}, \
  and\ \bibinfo {author} {\bibfnamefont {I.~B.}\ \bibnamefont {Spielman}},\
  }\bibfield  {title} {\emph {\bibinfo {title} {{Topological bands for
  ultracold atoms}},\ }}\href {\doibase 10.1103/RevModPhys.91.015005}
  {\bibfield  {journal} {\bibinfo  {journal} {Rev. Mod. Phys.}\ }\textbf
  {\bibinfo {volume} {91}},\ \bibinfo {pages} {015005} (\bibinfo {year}
  {2019})}\BibitemShut {NoStop}%
\bibitem [{\citenamefont {Ozawa}\ \emph {et~al.}(2019)\citenamefont {Ozawa},
  \citenamefont {Price}, \citenamefont {Amo}, \citenamefont {Goldman},
  \citenamefont {Hafezi}, \citenamefont {Lu}, \citenamefont {Rechtsman},
  \citenamefont {Schuster}, \citenamefont {Simon}, \citenamefont {Zilberberg},\
  and\ \citenamefont {Carusotto}}]{RevModPhys.91.015006}%
  \BibitemOpen
  \bibfield  {author} {\bibinfo {author} {\bibfnamefont {T.}~\bibnamefont
  {Ozawa}}, \bibinfo {author} {\bibfnamefont {H.~M.}\ \bibnamefont {Price}},
  \bibinfo {author} {\bibfnamefont {A.}~\bibnamefont {Amo}}, \bibinfo {author}
  {\bibfnamefont {N.}~\bibnamefont {Goldman}}, \bibinfo {author} {\bibfnamefont
  {M.}~\bibnamefont {Hafezi}}, \bibinfo {author} {\bibfnamefont
  {L.}~\bibnamefont {Lu}}, \bibinfo {author} {\bibfnamefont {M.~C.}\
  \bibnamefont {Rechtsman}}, \bibinfo {author} {\bibfnamefont {D.}~\bibnamefont
  {Schuster}}, \bibinfo {author} {\bibfnamefont {J.}~\bibnamefont {Simon}},
  \bibinfo {author} {\bibfnamefont {O.}~\bibnamefont {Zilberberg}}, \ and\
  \bibinfo {author} {\bibfnamefont {I.}~\bibnamefont {Carusotto}},\ }\bibfield
  {title} {\emph {\bibinfo {title} {Topological photonics},\ }}\href {\doibase
  10.1103/RevModPhys.91.015006} {\bibfield  {journal} {\bibinfo  {journal}
  {Rev. Mod. Phys.}\ }\textbf {\bibinfo {volume} {91}},\ \bibinfo {pages}
  {015006} (\bibinfo {year} {2019})}\BibitemShut {NoStop}%
\bibitem [{\citenamefont {Ohta}\ \emph {et~al.}(2016)\citenamefont {Ohta},
  \citenamefont {Tanaka}, \citenamefont {Danshita},\ and\ \citenamefont
  {Totsuka}}]{PhysRevB.93.165423}%
  \BibitemOpen
  \bibfield  {author} {\bibinfo {author} {\bibfnamefont {T.}~\bibnamefont
  {Ohta}}, \bibinfo {author} {\bibfnamefont {S.}~\bibnamefont {Tanaka}},
  \bibinfo {author} {\bibfnamefont {I.}~\bibnamefont {Danshita}}, \ and\
  \bibinfo {author} {\bibfnamefont {K.}~\bibnamefont {Totsuka}},\ }\bibfield
  {title} {\emph {\bibinfo {title} {{Topological and dynamical properties of a
  generalized cluster model in one dimension}},\ }}\href {\doibase
  10.1103/PhysRevB.93.165423} {\bibfield  {journal} {\bibinfo  {journal} {Phys.
  Rev. B}\ }\textbf {\bibinfo {volume} {93}},\ \bibinfo {pages} {165423}
  (\bibinfo {year} {2016})}\BibitemShut {NoStop}%
\bibitem [{\citenamefont {Nie}\ \emph {et~al.}(2017)\citenamefont {Nie},
  \citenamefont {Mei}, \citenamefont {Amico},\ and\ \citenamefont
  {Kwek}}]{PhysRevE.96.020106}%
  \BibitemOpen
  \bibfield  {author} {\bibinfo {author} {\bibfnamefont {W.}~\bibnamefont
  {Nie}}, \bibinfo {author} {\bibfnamefont {F.}~\bibnamefont {Mei}}, \bibinfo
  {author} {\bibfnamefont {L.}~\bibnamefont {Amico}}, \ and\ \bibinfo {author}
  {\bibfnamefont {L.~C.}\ \bibnamefont {Kwek}},\ }\bibfield  {title} {\emph
  {\bibinfo {title} {{Scaling of geometric phase versus band structure in
  cluster-Ising models}},\ }}\href {\doibase 10.1103/PhysRevE.96.020106}
  {\bibfield  {journal} {\bibinfo  {journal} {Phys. Rev. E}\ }\textbf {\bibinfo
  {volume} {96}},\ \bibinfo {pages} {020106(R)} (\bibinfo {year}
  {2017})}\BibitemShut {NoStop}%
\bibitem [{\citenamefont {Che}\ \emph {et~al.}(2020)\citenamefont {Che},
  \citenamefont {Gneiting}, \citenamefont {Liu},\ and\ \citenamefont
  {Nori}}]{PhysRevB.102.134213}%
  \BibitemOpen
  \bibfield  {author} {\bibinfo {author} {\bibfnamefont {Y.}~\bibnamefont
  {Che}}, \bibinfo {author} {\bibfnamefont {C.}~\bibnamefont {Gneiting}},
  \bibinfo {author} {\bibfnamefont {T.}~\bibnamefont {Liu}}, \ and\ \bibinfo
  {author} {\bibfnamefont {F.}~\bibnamefont {Nori}},\ }\bibfield  {title}
  {\emph {\bibinfo {title} {Topological quantum phase transitions retrieved
  through unsupervised machine learning},\ }}\href {\doibase
  10.1103/PhysRevB.102.134213} {\bibfield  {journal} {\bibinfo  {journal}
  {Phys. Rev. B}\ }\textbf {\bibinfo {volume} {102}},\ \bibinfo {pages}
  {134213} (\bibinfo {year} {2020})}\BibitemShut {NoStop}%
\bibitem [{\citenamefont {Bravyi}\ \emph {et~al.}(2010)\citenamefont {Bravyi},
  \citenamefont {Hastings},\ and\ \citenamefont
  {Michalakis}}]{bravyi2010topological}%
  \BibitemOpen
  \bibfield  {author} {\bibinfo {author} {\bibfnamefont {S.}~\bibnamefont
  {Bravyi}}, \bibinfo {author} {\bibfnamefont {M.~B.}\ \bibnamefont
  {Hastings}}, \ and\ \bibinfo {author} {\bibfnamefont {S.}~\bibnamefont
  {Michalakis}},\ }\bibfield  {title} {\emph {\bibinfo {title} {{Topological
  quantum order: stability under local perturbations}},\ }}\href {\doibase
  10.1063/1.3490195} {\bibfield  {journal} {\bibinfo  {journal} {J. Math.
  Phys.}\ }\textbf {\bibinfo {volume} {51}},\ \bibinfo {pages} {093512}
  (\bibinfo {year} {2010})}\BibitemShut {NoStop}%
\bibitem [{\citenamefont {Hafezi}\ \emph {et~al.}(2011)\citenamefont {Hafezi},
  \citenamefont {Demler}, \citenamefont {Lukin},\ and\ \citenamefont
  {Taylor}}]{hafezi2011robust}%
  \BibitemOpen
  \bibfield  {author} {\bibinfo {author} {\bibfnamefont {M.}~\bibnamefont
  {Hafezi}}, \bibinfo {author} {\bibfnamefont {E.~A.}\ \bibnamefont {Demler}},
  \bibinfo {author} {\bibfnamefont {M.~D.}\ \bibnamefont {Lukin}}, \ and\
  \bibinfo {author} {\bibfnamefont {J.~M.}\ \bibnamefont {Taylor}},\ }\bibfield
   {title} {\emph {\bibinfo {title} {{Robust optical delay lines with
  topological protection}},\ }}\href {\doibase 10.1038/nphys2063} {\bibfield
  {journal} {\bibinfo  {journal} {Nat. Phys.}\ }\textbf {\bibinfo {volume}
  {7}},\ \bibinfo {pages} {907} (\bibinfo {year} {2011})}\BibitemShut {NoStop}%
\bibitem [{\citenamefont {Cheng}\ \emph {et~al.}(2012)\citenamefont {Cheng},
  \citenamefont {Lutchyn},\ and\ \citenamefont
  {Das~Sarma}}]{PhysRevB.85.165124}%
  \BibitemOpen
  \bibfield  {author} {\bibinfo {author} {\bibfnamefont {M.}~\bibnamefont
  {Cheng}}, \bibinfo {author} {\bibfnamefont {R.~M.}\ \bibnamefont {Lutchyn}},
  \ and\ \bibinfo {author} {\bibfnamefont {S.}~\bibnamefont {Das~Sarma}},\
  }\bibfield  {title} {\emph {\bibinfo {title} {{Topological protection of
  Majorana qubits}},\ }}\href {\doibase 10.1103/PhysRevB.85.165124} {\bibfield
  {journal} {\bibinfo  {journal} {Phys. Rev. B}\ }\textbf {\bibinfo {volume}
  {85}},\ \bibinfo {pages} {165124} (\bibinfo {year} {2012})}\BibitemShut
  {NoStop}%
\bibitem [{\citenamefont {Mondragon-Shem}\ \emph {et~al.}(2014)\citenamefont
  {Mondragon-Shem}, \citenamefont {Hughes}, \citenamefont {Song},\ and\
  \citenamefont {Prodan}}]{PhysRevLett.113.046802}%
  \BibitemOpen
  \bibfield  {author} {\bibinfo {author} {\bibfnamefont {I.}~\bibnamefont
  {Mondragon-Shem}}, \bibinfo {author} {\bibfnamefont {T.~L.}\ \bibnamefont
  {Hughes}}, \bibinfo {author} {\bibfnamefont {J.}~\bibnamefont {Song}}, \ and\
  \bibinfo {author} {\bibfnamefont {E.}~\bibnamefont {Prodan}},\ }\bibfield
  {title} {\emph {\bibinfo {title} {{Topological Criticality in the
  Chiral-Symmetric AIII Class at Strong Disorder}},\ }}\href {\doibase
  10.1103/PhysRevLett.113.046802} {\bibfield  {journal} {\bibinfo  {journal}
  {Phys. Rev. Lett.}\ }\textbf {\bibinfo {volume} {113}},\ \bibinfo {pages}
  {046802} (\bibinfo {year} {2014})}\BibitemShut {NoStop}%
\bibitem [{\citenamefont {Poli}\ \emph {et~al.}(2015)\citenamefont {Poli},
  \citenamefont {Bellec}, \citenamefont {Kuhl}, \citenamefont {Mortessagne},\
  and\ \citenamefont {Schomerus}}]{poli2015selective}%
  \BibitemOpen
  \bibfield  {author} {\bibinfo {author} {\bibfnamefont {C.}~\bibnamefont
  {Poli}}, \bibinfo {author} {\bibfnamefont {M.}~\bibnamefont {Bellec}},
  \bibinfo {author} {\bibfnamefont {U.}~\bibnamefont {Kuhl}}, \bibinfo {author}
  {\bibfnamefont {F.}~\bibnamefont {Mortessagne}}, \ and\ \bibinfo {author}
  {\bibfnamefont {H.}~\bibnamefont {Schomerus}},\ }\bibfield  {title} {\emph
  {\bibinfo {title} {Selective enhancement of topologically induced interface
  states in a dielectric resonator chain},\ }}\href {\doibase
  10.1038/ncomms7710} {\bibfield  {journal} {\bibinfo  {journal} {Nat.
  Commun.}\ }\textbf {\bibinfo {volume} {6}},\ \bibinfo {pages} {6710}
  (\bibinfo {year} {2015})}\BibitemShut {NoStop}%
\bibitem [{\citenamefont {Balabanov}\ and\ \citenamefont
  {Johannesson}(2017)}]{PhysRevB.96.035149}%
  \BibitemOpen
  \bibfield  {author} {\bibinfo {author} {\bibfnamefont {O.}~\bibnamefont
  {Balabanov}}\ and\ \bibinfo {author} {\bibfnamefont {H.}~\bibnamefont
  {Johannesson}},\ }\bibfield  {title} {\emph {\bibinfo {title} {{Robustness of
  symmetry-protected topological states against time-periodic perturbations}},\
  }}\href {\doibase 10.1103/PhysRevB.96.035149} {\bibfield  {journal} {\bibinfo
   {journal} {Phys. Rev. B}\ }\textbf {\bibinfo {volume} {96}},\ \bibinfo
  {pages} {035149} (\bibinfo {year} {2017})}\BibitemShut {NoStop}%
\bibitem [{\citenamefont {Liu}\ \emph {et~al.}(2019)\citenamefont {Liu},
  \citenamefont {Zhang}, \citenamefont {Ai}, \citenamefont {Gong},
  \citenamefont {Kawabata}, \citenamefont {Ueda},\ and\ \citenamefont
  {Nori}}]{PhysRevLett.122.076801}%
  \BibitemOpen
  \bibfield  {author} {\bibinfo {author} {\bibfnamefont {T.}~\bibnamefont
  {Liu}}, \bibinfo {author} {\bibfnamefont {Y.-R.}\ \bibnamefont {Zhang}},
  \bibinfo {author} {\bibfnamefont {Q.}~\bibnamefont {Ai}}, \bibinfo {author}
  {\bibfnamefont {Z.}~\bibnamefont {Gong}}, \bibinfo {author} {\bibfnamefont
  {K.}~\bibnamefont {Kawabata}}, \bibinfo {author} {\bibfnamefont
  {M.}~\bibnamefont {Ueda}}, \ and\ \bibinfo {author} {\bibfnamefont
  {F.}~\bibnamefont {Nori}},\ }\bibfield  {title} {\emph {\bibinfo {title}
  {{Second-Order Topological Phases in Non-Hermitian Systems}},\ }}\href
  {\doibase 10.1103/PhysRevLett.122.076801} {\bibfield  {journal} {\bibinfo
  {journal} {Phys. Rev. Lett.}\ }\textbf {\bibinfo {volume} {122}},\ \bibinfo
  {pages} {076801} (\bibinfo {year} {2019})}\BibitemShut {NoStop}%
\bibitem [{\citenamefont {Wang}\ \emph {et~al.}(2019)\citenamefont {Wang},
  \citenamefont {Lu}, \citenamefont {Mei}, \citenamefont {Gao}, \citenamefont
  {Li}, \citenamefont {Tang}, \citenamefont {Zhu}, \citenamefont {Jia},\ and\
  \citenamefont {Jin}}]{PhysRevLett.122.193903}%
  \BibitemOpen
  \bibfield  {author} {\bibinfo {author} {\bibfnamefont {Y.}~\bibnamefont
  {Wang}}, \bibinfo {author} {\bibfnamefont {Y.-H.}\ \bibnamefont {Lu}},
  \bibinfo {author} {\bibfnamefont {F.}~\bibnamefont {Mei}}, \bibinfo {author}
  {\bibfnamefont {J.}~\bibnamefont {Gao}}, \bibinfo {author} {\bibfnamefont
  {Z.-M.}\ \bibnamefont {Li}}, \bibinfo {author} {\bibfnamefont
  {H.}~\bibnamefont {Tang}}, \bibinfo {author} {\bibfnamefont {S.-L.}\
  \bibnamefont {Zhu}}, \bibinfo {author} {\bibfnamefont {S.}~\bibnamefont
  {Jia}}, \ and\ \bibinfo {author} {\bibfnamefont {X.-M.}\ \bibnamefont
  {Jin}},\ }\bibfield  {title} {\emph {\bibinfo {title} {{Direct Observation of
  Topology from Single-Photon Dynamics}},\ }}\href {\doibase
  10.1103/PhysRevLett.122.193903} {\bibfield  {journal} {\bibinfo  {journal}
  {Phys. Rev. Lett.}\ }\textbf {\bibinfo {volume} {122}},\ \bibinfo {pages}
  {193903} (\bibinfo {year} {2019})}\BibitemShut {NoStop}%
\bibitem [{\citenamefont {Nie}\ \emph {et~al.}(2020)\citenamefont {Nie},
  \citenamefont {Peng}, \citenamefont {Nori},\ and\ \citenamefont
  {Liu}}]{PhysRevLett.124.023603}%
  \BibitemOpen
  \bibfield  {author} {\bibinfo {author} {\bibfnamefont {W.}~\bibnamefont
  {Nie}}, \bibinfo {author} {\bibfnamefont {Z.~H.}\ \bibnamefont {Peng}},
  \bibinfo {author} {\bibfnamefont {F.}~\bibnamefont {Nori}}, \ and\ \bibinfo
  {author} {\bibfnamefont {Y.-X.}\ \bibnamefont {Liu}},\ }\bibfield  {title}
  {\emph {\bibinfo {title} {{Topologically Protected Quantum Coherence in a
  Superatom}},\ }}\href {\doibase 10.1103/PhysRevLett.124.023603} {\bibfield
  {journal} {\bibinfo  {journal} {Phys. Rev. Lett.}\ }\textbf {\bibinfo
  {volume} {124}},\ \bibinfo {pages} {023603} (\bibinfo {year}
  {2020})}\BibitemShut {NoStop}%
\bibitem [{\citenamefont {Gong}\ \emph {et~al.}(2012)\citenamefont {Gong},
  \citenamefont {Chen}, \citenamefont {Jia},\ and\ \citenamefont
  {Zhang}}]{PhysRevLett.109.105302}%
  \BibitemOpen
  \bibfield  {author} {\bibinfo {author} {\bibfnamefont {M.}~\bibnamefont
  {Gong}}, \bibinfo {author} {\bibfnamefont {G.}~\bibnamefont {Chen}}, \bibinfo
  {author} {\bibfnamefont {S.}~\bibnamefont {Jia}}, \ and\ \bibinfo {author}
  {\bibfnamefont {C.}~\bibnamefont {Zhang}},\ }\bibfield  {title} {\emph
  {\bibinfo {title} {{Searching for Majorana Fermions in 2D Spin-Orbit Coupled
  Fermi Superfluids at Finite Temperature}},\ }}\href {\doibase
  10.1103/PhysRevLett.109.105302} {\bibfield  {journal} {\bibinfo  {journal}
  {Phys. Rev. Lett.}\ }\textbf {\bibinfo {volume} {109}},\ \bibinfo {pages}
  {105302} (\bibinfo {year} {2012})}\BibitemShut {NoStop}%
\bibitem [{\citenamefont {Viyuela}\ \emph {et~al.}(2014)\citenamefont
  {Viyuela}, \citenamefont {Rivas},\ and\ \citenamefont
  {Martin-Delgado}}]{PhysRevLett.112.130401}%
  \BibitemOpen
  \bibfield  {author} {\bibinfo {author} {\bibfnamefont {O.}~\bibnamefont
  {Viyuela}}, \bibinfo {author} {\bibfnamefont {A.}~\bibnamefont {Rivas}}, \
  and\ \bibinfo {author} {\bibfnamefont {M.~A.}\ \bibnamefont
  {Martin-Delgado}},\ }\bibfield  {title} {\emph {\bibinfo {title} {{Uhlmann
  Phase as a Topological Measure for One-Dimensional Fermion Systems}},\
  }}\href {\doibase 10.1103/PhysRevLett.112.130401} {\bibfield  {journal}
  {\bibinfo  {journal} {Phys. Rev. Lett.}\ }\textbf {\bibinfo {volume} {112}},\
  \bibinfo {pages} {130401} (\bibinfo {year} {2014})}\BibitemShut {NoStop}%
\bibitem [{\citenamefont {Huang}\ and\ \citenamefont
  {Arovas}(2014)}]{PhysRevLett.113.076407}%
  \BibitemOpen
  \bibfield  {author} {\bibinfo {author} {\bibfnamefont {Z.}~\bibnamefont
  {Huang}}\ and\ \bibinfo {author} {\bibfnamefont {D.~P.}\ \bibnamefont
  {Arovas}},\ }\bibfield  {title} {\emph {\bibinfo {title} {{Topological
  Indices for Open and Thermal Systems Via Uhlmann's Phase}},\ }}\href
  {\doibase 10.1103/PhysRevLett.113.076407} {\bibfield  {journal} {\bibinfo
  {journal} {Phys. Rev. Lett.}\ }\textbf {\bibinfo {volume} {113}},\ \bibinfo
  {pages} {076407} (\bibinfo {year} {2014})}\BibitemShut {NoStop}%
\bibitem [{\citenamefont {Monserrat}\ and\ \citenamefont
  {Vanderbilt}(2016)}]{PhysRevLett.117.226801}%
  \BibitemOpen
  \bibfield  {author} {\bibinfo {author} {\bibfnamefont {B.}~\bibnamefont
  {Monserrat}}\ and\ \bibinfo {author} {\bibfnamefont {D.}~\bibnamefont
  {Vanderbilt}},\ }\bibfield  {title} {\emph {\bibinfo {title} {{Temperature
  Effects in the Band Structure of Topological Insulators}},\ }}\href {\doibase
  10.1103/PhysRevLett.117.226801} {\bibfield  {journal} {\bibinfo  {journal}
  {Phys. Rev. Lett.}\ }\textbf {\bibinfo {volume} {117}},\ \bibinfo {pages}
  {226801} (\bibinfo {year} {2016})}\BibitemShut {NoStop}%
\bibitem [{\citenamefont {Bardyn}\ \emph {et~al.}(2018)\citenamefont {Bardyn},
  \citenamefont {Wawer}, \citenamefont {Altland}, \citenamefont
  {Fleischhauer},\ and\ \citenamefont {Diehl}}]{PhysRevX.8.011035}%
  \BibitemOpen
  \bibfield  {author} {\bibinfo {author} {\bibfnamefont {C.-E.}\ \bibnamefont
  {Bardyn}}, \bibinfo {author} {\bibfnamefont {L.}~\bibnamefont {Wawer}},
  \bibinfo {author} {\bibfnamefont {A.}~\bibnamefont {Altland}}, \bibinfo
  {author} {\bibfnamefont {M.}~\bibnamefont {Fleischhauer}}, \ and\ \bibinfo
  {author} {\bibfnamefont {S.}~\bibnamefont {Diehl}},\ }\bibfield  {title}
  {\emph {\bibinfo {title} {{Probing the Topology of Density Matrices}},\
  }}\href {\doibase 10.1103/PhysRevX.8.011035} {\bibfield  {journal} {\bibinfo
  {journal} {Phys. Rev. X}\ }\textbf {\bibinfo {volume} {8}},\ \bibinfo {pages}
  {011035} (\bibinfo {year} {2018})}\BibitemShut {NoStop}%
\bibitem [{\citenamefont {Unanyan}\ \emph {et~al.}(2020)\citenamefont
  {Unanyan}, \citenamefont {Kiefer-Emmanouilidis},\ and\ \citenamefont
  {Fleischhauer}}]{PhysRevLett.125.215701}%
  \BibitemOpen
  \bibfield  {author} {\bibinfo {author} {\bibfnamefont {R.}~\bibnamefont
  {Unanyan}}, \bibinfo {author} {\bibfnamefont {M.}~\bibnamefont
  {Kiefer-Emmanouilidis}}, \ and\ \bibinfo {author} {\bibfnamefont
  {M.}~\bibnamefont {Fleischhauer}},\ }\bibfield  {title} {\emph {\bibinfo
  {title} {{Finite-Temperature Topological Invariant for Interacting
  Systems}},\ }}\href {\doibase 10.1103/PhysRevLett.125.215701} {\bibfield
  {journal} {\bibinfo  {journal} {Phys. Rev. Lett.}\ }\textbf {\bibinfo
  {volume} {125}},\ \bibinfo {pages} {215701} (\bibinfo {year}
  {2020})}\BibitemShut {NoStop}%
\bibitem [{\citenamefont {McGinley}\ and\ \citenamefont
  {Cooper}(2020)}]{McGinley2020fragility}%
  \BibitemOpen
  \bibfield  {author} {\bibinfo {author} {\bibfnamefont {M.}~\bibnamefont
  {McGinley}}\ and\ \bibinfo {author} {\bibfnamefont {N.~R.}\ \bibnamefont
  {Cooper}},\ }\bibfield  {title} {\emph {\bibinfo {title} {{Fragility of
  time-reversal symmetry protected topological phases}},\ }}\href {\doibase
  10.1038/s41567-020-0956-z} {\bibfield  {journal} {\bibinfo  {journal} {Nat.
  Phys.}\ }\textbf {\bibinfo {volume} {16}},\ \bibinfo {pages} {1181} (\bibinfo
  {year} {2020})}\BibitemShut {NoStop}%
\bibitem [{\citenamefont {Breuer}\ and\ \citenamefont
  {Petruccione}(2002)}]{breuer2002theory}%
  \BibitemOpen
  \bibfield  {author} {\bibinfo {author} {\bibfnamefont {H.-P.}\ \bibnamefont
  {Breuer}}\ and\ \bibinfo {author} {\bibfnamefont {F.}~\bibnamefont
  {Petruccione}},\ }\href@noop {} {\emph {\bibinfo {title} {{The theory of open
  quantum systems}}}}\ (\bibinfo  {publisher} {Oxford University Press, New
  York},\ \bibinfo {year} {2002})\BibitemShut {NoStop}%
\bibitem [{Sup()}]{SupplementalMaterial}%
  \BibitemOpen
  \href@noop {} {\ }\bibinfo {note} {{See the Supplemental Material at
  http://link.aps.org/ supplemental/10.1103/PhysRevLett.127.250402 for
  additional details about the derivation of the master equation,
  environment-induced topological phase transition, topological protection of
  quantum coherence, and experimental scheme for detecting the dissipative
  topological phase transition.}}\BibitemShut {Stop}%
\bibitem [{\citenamefont {Chen}\ \emph {et~al.}(2014)\citenamefont {Chen},
  \citenamefont {Neill}, \citenamefont {Roushan}, \citenamefont {Leung},
  \citenamefont {Fang}, \citenamefont {Barends}, \citenamefont {Kelly},
  \citenamefont {Campbell}, \citenamefont {Chen}, \citenamefont {Chiaro},
  \citenamefont {Dunsworth}, \citenamefont {Jeffrey}, \citenamefont {Megrant},
  \citenamefont {Mutus}, \citenamefont {O'Malley}, \citenamefont {Quintana},
  \citenamefont {Sank}, \citenamefont {Vainsencher}, \citenamefont {Wenner},
  \citenamefont {White}, \citenamefont {Geller}, \citenamefont {Cleland},\ and\
  \citenamefont {Martinis}}]{PhysRevLett.113.220502}%
  \BibitemOpen
  \bibfield  {author} {\bibinfo {author} {\bibfnamefont {Y.}~\bibnamefont
  {Chen}}, \bibinfo {author} {\bibfnamefont {C.}~\bibnamefont {Neill}},
  \bibinfo {author} {\bibfnamefont {P.}~\bibnamefont {Roushan}}, \bibinfo
  {author} {\bibfnamefont {N.}~\bibnamefont {Leung}}, \bibinfo {author}
  {\bibfnamefont {M.}~\bibnamefont {Fang}}, \bibinfo {author} {\bibfnamefont
  {R.}~\bibnamefont {Barends}}, \bibinfo {author} {\bibfnamefont
  {J.}~\bibnamefont {Kelly}}, \bibinfo {author} {\bibfnamefont
  {B.}~\bibnamefont {Campbell}}, \bibinfo {author} {\bibfnamefont
  {Z.}~\bibnamefont {Chen}}, \bibinfo {author} {\bibfnamefont {B.}~\bibnamefont
  {Chiaro}}, \bibinfo {author} {\bibfnamefont {A.}~\bibnamefont {Dunsworth}},
  \bibinfo {author} {\bibfnamefont {E.}~\bibnamefont {Jeffrey}}, \bibinfo
  {author} {\bibfnamefont {A.}~\bibnamefont {Megrant}}, \bibinfo {author}
  {\bibfnamefont {J.~Y.}\ \bibnamefont {Mutus}}, \bibinfo {author}
  {\bibfnamefont {P.~J.~J.}\ \bibnamefont {O'Malley}}, \bibinfo {author}
  {\bibfnamefont {C.~M.}\ \bibnamefont {Quintana}}, \bibinfo {author}
  {\bibfnamefont {D.}~\bibnamefont {Sank}}, \bibinfo {author} {\bibfnamefont
  {A.}~\bibnamefont {Vainsencher}}, \bibinfo {author} {\bibfnamefont
  {J.}~\bibnamefont {Wenner}}, \bibinfo {author} {\bibfnamefont {T.~C.}\
  \bibnamefont {White}}, \bibinfo {author} {\bibfnamefont {M.~R.}\ \bibnamefont
  {Geller}}, \bibinfo {author} {\bibfnamefont {A.~N.}\ \bibnamefont {Cleland}},
  \ and\ \bibinfo {author} {\bibfnamefont {J.~M.}\ \bibnamefont {Martinis}},\
  }\bibfield  {title} {\emph {\bibinfo {title} {{Qubit Architecture with High
  Coherence and Fast Tunable Coupling}},\ }}\href {\doibase
  10.1103/PhysRevLett.113.220502} {\bibfield  {journal} {\bibinfo  {journal}
  {Phys. Rev. Lett.}\ }\textbf {\bibinfo {volume} {113}},\ \bibinfo {pages}
  {220502} (\bibinfo {year} {2014})}\BibitemShut {NoStop}%
\bibitem [{\citenamefont {Knoll}\ \emph {et~al.}(2000)\citenamefont {Knoll},
  \citenamefont {Scheel},\ and\ \citenamefont {Welsch}}]{knoll2000qed}%
  \BibitemOpen
  \bibfield  {author} {\bibinfo {author} {\bibfnamefont {L.}~\bibnamefont
  {Knoll}}, \bibinfo {author} {\bibfnamefont {S.}~\bibnamefont {Scheel}}, \
  and\ \bibinfo {author} {\bibfnamefont {D.-G.}\ \bibnamefont {Welsch}},\
  }\bibfield  {title} {\emph {\bibinfo {title} {{QED in dispersing and
  absorbing media}},\ }}\href@noop {} {\bibfield  {journal} {\bibinfo
  {journal} {arXiv:quant-ph/0006121}\ } (\bibinfo {year} {2000})}\BibitemShut
  {NoStop}%
\bibitem [{\citenamefont {Dung}\ \emph {et~al.}(2002)\citenamefont {Dung},
  \citenamefont {Kn\"oll},\ and\ \citenamefont {Welsch}}]{PhysRevA.66.063810}%
  \BibitemOpen
  \bibfield  {author} {\bibinfo {author} {\bibfnamefont {H.~T.}\ \bibnamefont
  {Dung}}, \bibinfo {author} {\bibfnamefont {L.}~\bibnamefont {Kn\"oll}}, \
  and\ \bibinfo {author} {\bibfnamefont {D.-G.}\ \bibnamefont {Welsch}},\
  }\bibfield  {title} {\emph {\bibinfo {title} {{Resonant dipole-dipole
  interaction in the presence of dispersing and absorbing surroundings}},\
  }}\href {\doibase 10.1103/PhysRevA.66.063810} {\bibfield  {journal} {\bibinfo
   {journal} {Phys. Rev. A}\ }\textbf {\bibinfo {volume} {66}},\ \bibinfo
  {pages} {063810} (\bibinfo {year} {2002})}\BibitemShut {NoStop}%
\bibitem [{\citenamefont {Dzsotjan}\ \emph {et~al.}(2010)\citenamefont
  {Dzsotjan}, \citenamefont {S\o{}rensen},\ and\ \citenamefont
  {Fleischhauer}}]{PhysRevB.82.075427}%
  \BibitemOpen
  \bibfield  {author} {\bibinfo {author} {\bibfnamefont {D.}~\bibnamefont
  {Dzsotjan}}, \bibinfo {author} {\bibfnamefont {A.~S.}\ \bibnamefont
  {S\o{}rensen}}, \ and\ \bibinfo {author} {\bibfnamefont {M.}~\bibnamefont
  {Fleischhauer}},\ }\bibfield  {title} {\emph {\bibinfo {title} {{Quantum
  emitters coupled to surface plasmons of a nanowire: A Green's function
  approach}},\ }}\href {\doibase 10.1103/PhysRevB.82.075427} {\bibfield
  {journal} {\bibinfo  {journal} {Phys. Rev. B}\ }\textbf {\bibinfo {volume}
  {82}},\ \bibinfo {pages} {075427} (\bibinfo {year} {2010})}\BibitemShut
  {NoStop}%
\bibitem [{\citenamefont {Gonzalez-Tudela}\ \emph {et~al.}(2011)\citenamefont
  {Gonzalez-Tudela}, \citenamefont {Martin-Cano}, \citenamefont {Moreno},
  \citenamefont {Martin-Moreno}, \citenamefont {Tejedor},\ and\ \citenamefont
  {Garcia-Vidal}}]{PhysRevLett.106.020501}%
  \BibitemOpen
  \bibfield  {author} {\bibinfo {author} {\bibfnamefont {A.}~\bibnamefont
  {Gonzalez-Tudela}}, \bibinfo {author} {\bibfnamefont {D.}~\bibnamefont
  {Martin-Cano}}, \bibinfo {author} {\bibfnamefont {E.}~\bibnamefont {Moreno}},
  \bibinfo {author} {\bibfnamefont {L.}~\bibnamefont {Martin-Moreno}}, \bibinfo
  {author} {\bibfnamefont {C.}~\bibnamefont {Tejedor}}, \ and\ \bibinfo
  {author} {\bibfnamefont {F.~J.}\ \bibnamefont {Garcia-Vidal}},\ }\bibfield
  {title} {\emph {\bibinfo {title} {{Entanglement of Two Qubits Mediated by
  One-Dimensional Plasmonic Waveguides}},\ }}\href {\doibase
  10.1103/PhysRevLett.106.020501} {\bibfield  {journal} {\bibinfo  {journal}
  {Phys. Rev. Lett.}\ }\textbf {\bibinfo {volume} {106}},\ \bibinfo {pages}
  {020501} (\bibinfo {year} {2011})}\BibitemShut {NoStop}%
\bibitem [{\citenamefont {Mart\'{\i}n-Cano}\ \emph {et~al.}(2011)\citenamefont
  {Mart\'{\i}n-Cano}, \citenamefont {Gonz\'alez-Tudela}, \citenamefont
  {Mart\'{\i}n-Moreno}, \citenamefont {Garc\'{\i}a-Vidal}, \citenamefont
  {Tejedor},\ and\ \citenamefont {Moreno}}]{PhysRevB.84.235306}%
  \BibitemOpen
  \bibfield  {author} {\bibinfo {author} {\bibfnamefont {D.}~\bibnamefont
  {Mart\'{\i}n-Cano}}, \bibinfo {author} {\bibfnamefont {A.}~\bibnamefont
  {Gonz\'alez-Tudela}}, \bibinfo {author} {\bibfnamefont {L.}~\bibnamefont
  {Mart\'{\i}n-Moreno}}, \bibinfo {author} {\bibfnamefont {F.~J.}\ \bibnamefont
  {Garc\'{\i}a-Vidal}}, \bibinfo {author} {\bibfnamefont {C.}~\bibnamefont
  {Tejedor}}, \ and\ \bibinfo {author} {\bibfnamefont {E.}~\bibnamefont
  {Moreno}},\ }\bibfield  {title} {\emph {\bibinfo {title} {{Dissipation-driven
  generation of two-qubit entanglement mediated by plasmonic waveguides}},\
  }}\href {\doibase 10.1103/PhysRevB.84.235306} {\bibfield  {journal} {\bibinfo
   {journal} {Phys. Rev. B}\ }\textbf {\bibinfo {volume} {84}},\ \bibinfo
  {pages} {235306} (\bibinfo {year} {2011})}\BibitemShut {NoStop}%
\bibitem [{\citenamefont {Angelatos}\ and\ \citenamefont
  {Hughes}(2015)}]{PhysRevA.91.051803}%
  \BibitemOpen
  \bibfield  {author} {\bibinfo {author} {\bibfnamefont {G.}~\bibnamefont
  {Angelatos}}\ and\ \bibinfo {author} {\bibfnamefont {S.}~\bibnamefont
  {Hughes}},\ }\bibfield  {title} {\emph {\bibinfo {title} {{Entanglement
  dynamics and Mollow nonuplets between two coupled quantum dots in a nanowire
  photonic-crystal system}},\ }}\href {\doibase 10.1103/PhysRevA.91.051803}
  {\bibfield  {journal} {\bibinfo  {journal} {Phys. Rev. A}\ }\textbf {\bibinfo
  {volume} {91}},\ \bibinfo {pages} {051803} (\bibinfo {year}
  {2015})}\BibitemShut {NoStop}%
\bibitem [{\citenamefont {Asenjo-Garcia}\ \emph
  {et~al.}(2017{\natexlab{a}})\citenamefont {Asenjo-Garcia}, \citenamefont
  {Hood}, \citenamefont {Chang},\ and\ \citenamefont
  {Kimble}}]{PhysRevA.95.033818}%
  \BibitemOpen
  \bibfield  {author} {\bibinfo {author} {\bibfnamefont {A.}~\bibnamefont
  {Asenjo-Garcia}}, \bibinfo {author} {\bibfnamefont {J.~D.}\ \bibnamefont
  {Hood}}, \bibinfo {author} {\bibfnamefont {D.~E.}\ \bibnamefont {Chang}}, \
  and\ \bibinfo {author} {\bibfnamefont {H.~J.}\ \bibnamefont {Kimble}},\
  }\bibfield  {title} {\emph {\bibinfo {title} {{Atom-light interactions in
  quasi-one-dimensional nanostructures: A Green's-function perspective}},\
  }}\href {\doibase 10.1103/PhysRevA.95.033818} {\bibfield  {journal} {\bibinfo
   {journal} {Phys. Rev. A}\ }\textbf {\bibinfo {volume} {95}},\ \bibinfo
  {pages} {033818} (\bibinfo {year} {2017}{\natexlab{a}})}\BibitemShut
  {NoStop}%
\bibitem [{\citenamefont {Gangaraj}\ \emph {et~al.}(2017)\citenamefont
  {Gangaraj}, \citenamefont {Hanson},\ and\ \citenamefont
  {Antezza}}]{PhysRevA.95.063807}%
  \BibitemOpen
  \bibfield  {author} {\bibinfo {author} {\bibfnamefont {S.~A.~H.}\
  \bibnamefont {Gangaraj}}, \bibinfo {author} {\bibfnamefont {G.~W.}\
  \bibnamefont {Hanson}}, \ and\ \bibinfo {author} {\bibfnamefont
  {M.}~\bibnamefont {Antezza}},\ }\bibfield  {title} {\emph {\bibinfo {title}
  {{Robust entanglement with three-dimensional nonreciprocal photonic
  topological insulators}},\ }}\href {\doibase 10.1103/PhysRevA.95.063807}
  {\bibfield  {journal} {\bibinfo  {journal} {Phys. Rev. A}\ }\textbf {\bibinfo
  {volume} {95}},\ \bibinfo {pages} {063807} (\bibinfo {year}
  {2017})}\BibitemShut {NoStop}%
\bibitem [{\citenamefont {Asenjo-Garcia}\ \emph
  {et~al.}(2017{\natexlab{b}})\citenamefont {Asenjo-Garcia}, \citenamefont
  {Moreno-Cardoner}, \citenamefont {Albrecht}, \citenamefont {Kimble},\ and\
  \citenamefont {Chang}}]{PhysRevX.7.031024}%
  \BibitemOpen
  \bibfield  {author} {\bibinfo {author} {\bibfnamefont {A.}~\bibnamefont
  {Asenjo-Garcia}}, \bibinfo {author} {\bibfnamefont {M.}~\bibnamefont
  {Moreno-Cardoner}}, \bibinfo {author} {\bibfnamefont {A.}~\bibnamefont
  {Albrecht}}, \bibinfo {author} {\bibfnamefont {H.~J.}\ \bibnamefont
  {Kimble}}, \ and\ \bibinfo {author} {\bibfnamefont {D.~E.}\ \bibnamefont
  {Chang}},\ }\bibfield  {title} {\emph {\bibinfo {title} {{Exponential
  Improvement in Photon Storage Fidelities Using Subradiance and ``Selective
  Radiance'' in Atomic Arrays}},\ }}\href {\doibase 10.1103/PhysRevX.7.031024}
  {\bibfield  {journal} {\bibinfo  {journal} {Phys. Rev. X}\ }\textbf {\bibinfo
  {volume} {7}},\ \bibinfo {pages} {031024} (\bibinfo {year}
  {2017}{\natexlab{b}})}\BibitemShut {NoStop}%
\bibitem [{\citenamefont {Doyeux}\ \emph {et~al.}(2017)\citenamefont {Doyeux},
  \citenamefont {Gangaraj}, \citenamefont {Hanson},\ and\ \citenamefont
  {Antezza}}]{PhysRevLett.119.173901}%
  \BibitemOpen
  \bibfield  {author} {\bibinfo {author} {\bibfnamefont {P.}~\bibnamefont
  {Doyeux}}, \bibinfo {author} {\bibfnamefont {S.~A.~H.}\ \bibnamefont
  {Gangaraj}}, \bibinfo {author} {\bibfnamefont {G.~W.}\ \bibnamefont
  {Hanson}}, \ and\ \bibinfo {author} {\bibfnamefont {M.}~\bibnamefont
  {Antezza}},\ }\bibfield  {title} {\emph {\bibinfo {title} {{Giant Interatomic
  Energy-Transport Amplification with Nonreciprocal Photonic Topological
  Insulators}},\ }}\href {\doibase 10.1103/PhysRevLett.119.173901} {\bibfield
  {journal} {\bibinfo  {journal} {Phys. Rev. Lett.}\ }\textbf {\bibinfo
  {volume} {119}},\ \bibinfo {pages} {173901} (\bibinfo {year}
  {2017})}\BibitemShut {NoStop}%
\bibitem [{\citenamefont {Shi}\ \emph {et~al.}(2015)\citenamefont {Shi},
  \citenamefont {Chang},\ and\ \citenamefont {Cirac}}]{PhysRevA.92.053834}%
  \BibitemOpen
  \bibfield  {author} {\bibinfo {author} {\bibfnamefont {T.}~\bibnamefont
  {Shi}}, \bibinfo {author} {\bibfnamefont {D.~E.}\ \bibnamefont {Chang}}, \
  and\ \bibinfo {author} {\bibfnamefont {J.~I.}\ \bibnamefont {Cirac}},\
  }\bibfield  {title} {\emph {\bibinfo {title} {Multiphoton-scattering theory
  and generalized master equations},\ }}\href {\doibase
  10.1103/PhysRevA.92.053834} {\bibfield  {journal} {\bibinfo  {journal} {Phys.
  Rev. A}\ }\textbf {\bibinfo {volume} {92}},\ \bibinfo {pages} {053834}
  (\bibinfo {year} {2015})}\BibitemShut {NoStop}%
\bibitem [{\citenamefont {Calaj\'o}\ \emph {et~al.}(2016)\citenamefont
  {Calaj\'o}, \citenamefont {Ciccarello}, \citenamefont {Chang},\ and\
  \citenamefont {Rabl}}]{PhysRevA.93.033833}%
  \BibitemOpen
  \bibfield  {author} {\bibinfo {author} {\bibfnamefont {G.}~\bibnamefont
  {Calaj\'o}}, \bibinfo {author} {\bibfnamefont {F.}~\bibnamefont
  {Ciccarello}}, \bibinfo {author} {\bibfnamefont {D.}~\bibnamefont {Chang}}, \
  and\ \bibinfo {author} {\bibfnamefont {P.}~\bibnamefont {Rabl}},\ }\bibfield
  {title} {\emph {\bibinfo {title} {{Atom-field dressed states in slow-light
  waveguide QED}},\ }}\href {\doibase 10.1103/PhysRevA.93.033833} {\bibfield
  {journal} {\bibinfo  {journal} {Phys. Rev. A}\ }\textbf {\bibinfo {volume}
  {93}},\ \bibinfo {pages} {033833} (\bibinfo {year} {2016})}\BibitemShut
  {NoStop}%
\bibitem [{\citenamefont {Mirhosseini}\ \emph {et~al.}(2019)\citenamefont
  {Mirhosseini}, \citenamefont {Kim}, \citenamefont {Zhang}, \citenamefont
  {Sipahigil}, \citenamefont {Dieterle}, \citenamefont {Keller}, \citenamefont
  {Asenjo-Garcia}, \citenamefont {Chang},\ and\ \citenamefont
  {Painter}}]{mirhosseini2019cavity}%
  \BibitemOpen
  \bibfield  {author} {\bibinfo {author} {\bibfnamefont {M.}~\bibnamefont
  {Mirhosseini}}, \bibinfo {author} {\bibfnamefont {E.}~\bibnamefont {Kim}},
  \bibinfo {author} {\bibfnamefont {X.}~\bibnamefont {Zhang}}, \bibinfo
  {author} {\bibfnamefont {A.}~\bibnamefont {Sipahigil}}, \bibinfo {author}
  {\bibfnamefont {P.~B.}\ \bibnamefont {Dieterle}}, \bibinfo {author}
  {\bibfnamefont {A.~J.}\ \bibnamefont {Keller}}, \bibinfo {author}
  {\bibfnamefont {A.}~\bibnamefont {Asenjo-Garcia}}, \bibinfo {author}
  {\bibfnamefont {D.~E.}\ \bibnamefont {Chang}}, \ and\ \bibinfo {author}
  {\bibfnamefont {O.}~\bibnamefont {Painter}},\ }\bibfield  {title} {\emph
  {\bibinfo {title} {{Cavity quantum electrodynamics with atom-like mirrors}},\
  }}\href {\doibase 10.1038/s41586-019-1196-1} {\bibfield  {journal} {\bibinfo
  {journal} {Nature}\ }\textbf {\bibinfo {volume} {569}},\ \bibinfo {pages}
  {692} (\bibinfo {year} {2019})}\BibitemShut {NoStop}%
\bibitem [{\citenamefont {Wen}\ \emph {et~al.}(2019)\citenamefont {Wen},
  \citenamefont {Lin}, \citenamefont {Kockum}, \citenamefont {Suri},
  \citenamefont {Ian}, \citenamefont {Chen}, \citenamefont {Mao}, \citenamefont
  {Chiu}, \citenamefont {Delsing}, \citenamefont {Nori}, \citenamefont {Lin},\
  and\ \citenamefont {Hoi}}]{PhysRevLett.123.233602}%
  \BibitemOpen
  \bibfield  {author} {\bibinfo {author} {\bibfnamefont {P.~Y.}\ \bibnamefont
  {Wen}}, \bibinfo {author} {\bibfnamefont {K.-T.}\ \bibnamefont {Lin}},
  \bibinfo {author} {\bibfnamefont {A.~F.}\ \bibnamefont {Kockum}}, \bibinfo
  {author} {\bibfnamefont {B.}~\bibnamefont {Suri}}, \bibinfo {author}
  {\bibfnamefont {H.}~\bibnamefont {Ian}}, \bibinfo {author} {\bibfnamefont
  {J.~C.}\ \bibnamefont {Chen}}, \bibinfo {author} {\bibfnamefont {S.~Y.}\
  \bibnamefont {Mao}}, \bibinfo {author} {\bibfnamefont {C.~C.}\ \bibnamefont
  {Chiu}}, \bibinfo {author} {\bibfnamefont {P.}~\bibnamefont {Delsing}},
  \bibinfo {author} {\bibfnamefont {F.}~\bibnamefont {Nori}}, \bibinfo {author}
  {\bibfnamefont {G.-D.}\ \bibnamefont {Lin}}, \ and\ \bibinfo {author}
  {\bibfnamefont {I.-C.}\ \bibnamefont {Hoi}},\ }\bibfield  {title} {\emph
  {\bibinfo {title} {{Large Collective Lamb Shift of Two Distant
  Superconducting Artificial Atoms}},\ }}\href {\doibase
  10.1103/PhysRevLett.123.233602} {\bibfield  {journal} {\bibinfo  {journal}
  {Phys. Rev. Lett.}\ }\textbf {\bibinfo {volume} {123}},\ \bibinfo {pages}
  {233602} (\bibinfo {year} {2019})}\BibitemShut {NoStop}%
\bibitem [{\citenamefont {Zanner}\ \emph {et~al.}(2021)\citenamefont {Zanner},
  \citenamefont {Orell}, \citenamefont {Schneider}, \citenamefont {Albert},
  \citenamefont {Oleschko}, \citenamefont {Juan}, \citenamefont {Silveri},\
  and\ \citenamefont {Kirchmair}}]{zanner2021coherent}%
  \BibitemOpen
  \bibfield  {author} {\bibinfo {author} {\bibfnamefont {M.}~\bibnamefont
  {Zanner}}, \bibinfo {author} {\bibfnamefont {T.}~\bibnamefont {Orell}},
  \bibinfo {author} {\bibfnamefont {C.~M.}\ \bibnamefont {Schneider}}, \bibinfo
  {author} {\bibfnamefont {R.}~\bibnamefont {Albert}}, \bibinfo {author}
  {\bibfnamefont {S.}~\bibnamefont {Oleschko}}, \bibinfo {author}
  {\bibfnamefont {M.~L.}\ \bibnamefont {Juan}}, \bibinfo {author}
  {\bibfnamefont {M.}~\bibnamefont {Silveri}}, \ and\ \bibinfo {author}
  {\bibfnamefont {G.}~\bibnamefont {Kirchmair}},\ }\bibfield  {title} {\emph
  {\bibinfo {title} {Coherent control of a symmetry-engineered multi-qubit dark
  state in waveguide quantum electrodynamics},\ }}\href
  {https://arxiv.org/abs/2106.05623} {\bibfield  {journal} {\bibinfo  {journal}
  {arXiv:2106.05623}\ } (\bibinfo {year} {2021})}\BibitemShut {NoStop}%
\bibitem [{\citenamefont {Su}\ \emph {et~al.}(1979)\citenamefont {Su},
  \citenamefont {Schrieffer},\ and\ \citenamefont
  {Heeger}}]{PhysRevLett.42.1698}%
  \BibitemOpen
  \bibfield  {author} {\bibinfo {author} {\bibfnamefont {W.~P.}\ \bibnamefont
  {Su}}, \bibinfo {author} {\bibfnamefont {J.~R.}\ \bibnamefont {Schrieffer}},
  \ and\ \bibinfo {author} {\bibfnamefont {A.~J.}\ \bibnamefont {Heeger}},\
  }\bibfield  {title} {\emph {\bibinfo {title} {{Solitons in Polyacetylene}},\
  }}\href {\doibase 10.1103/PhysRevLett.42.1698} {\bibfield  {journal}
  {\bibinfo  {journal} {Phys. Rev. Lett.}\ }\textbf {\bibinfo {volume} {42}},\
  \bibinfo {pages} {1698} (\bibinfo {year} {1979})}\BibitemShut {NoStop}%
\bibitem [{\citenamefont {Asb{\'o}th}\ \emph {et~al.}(2016)\citenamefont
  {Asb{\'o}th}, \citenamefont {Oroszl{\'a}ny},\ and\ \citenamefont
  {P{\'a}lyi}}]{asboth2016short}%
  \BibitemOpen
  \bibfield  {author} {\bibinfo {author} {\bibfnamefont {J.~K.}\ \bibnamefont
  {Asb{\'o}th}}, \bibinfo {author} {\bibfnamefont {L.}~\bibnamefont
  {Oroszl{\'a}ny}}, \ and\ \bibinfo {author} {\bibfnamefont {A.}~\bibnamefont
  {P{\'a}lyi}},\ }\href {\doibase 10.1007%2F978-3-319-25607-8} {\emph {\bibinfo
  {title} {{A Short Course on Topological Insulators}}}}\ (\bibinfo
  {publisher} {Springer},\ \bibinfo {year} {2016})\BibitemShut {NoStop}%
\bibitem [{\citenamefont {Murphy}\ \emph {et~al.}(2011)\citenamefont {Murphy},
  \citenamefont {Wortis},\ and\ \citenamefont {Atkinson}}]{PhysRevB.83.184206}%
  \BibitemOpen
  \bibfield  {author} {\bibinfo {author} {\bibfnamefont {N.~C.}\ \bibnamefont
  {Murphy}}, \bibinfo {author} {\bibfnamefont {R.}~\bibnamefont {Wortis}}, \
  and\ \bibinfo {author} {\bibfnamefont {W.~A.}\ \bibnamefont {Atkinson}},\
  }\bibfield  {title} {\emph {\bibinfo {title} {{Generalized inverse
  participation ratio as a possible measure of localization for interacting
  systems}},\ }}\href {\doibase 10.1103/PhysRevB.83.184206} {\bibfield
  {journal} {\bibinfo  {journal} {Phys. Rev. B}\ }\textbf {\bibinfo {volume}
  {83}},\ \bibinfo {pages} {184206} (\bibinfo {year} {2011})}\BibitemShut
  {NoStop}%
\bibitem [{\citenamefont {Nie}\ \emph {et~al.}(2021)\citenamefont {Nie},
  \citenamefont {Shi}, \citenamefont {Nori},\ and\ \citenamefont
  {Liu}}]{nie2020topology}%
  \BibitemOpen
  \bibfield  {author} {\bibinfo {author} {\bibfnamefont {W.}~\bibnamefont
  {Nie}}, \bibinfo {author} {\bibfnamefont {T.}~\bibnamefont {Shi}}, \bibinfo
  {author} {\bibfnamefont {F.}~\bibnamefont {Nori}}, \ and\ \bibinfo {author}
  {\bibfnamefont {Y.-X.}\ \bibnamefont {Liu}},\ }\bibfield  {title} {\emph
  {\bibinfo {title} {{Topology-Enhanced Nonreciprocal Scattering and Photon
  Absorption in a Waveguide}},\ }}\href {\doibase
  10.1103/PhysRevApplied.15.044041} {\bibfield  {journal} {\bibinfo  {journal}
  {Phys. Rev. Applied}\ }\textbf {\bibinfo {volume} {15}},\ \bibinfo {pages}
  {044041} (\bibinfo {year} {2021})}\BibitemShut {NoStop}%
\bibitem [{\citenamefont {Albert}\ \emph {et~al.}(2016)\citenamefont {Albert},
  \citenamefont {Bradlyn}, \citenamefont {Fraas},\ and\ \citenamefont
  {Jiang}}]{PhysRevX.6.041031}%
  \BibitemOpen
  \bibfield  {author} {\bibinfo {author} {\bibfnamefont {V.~V.}\ \bibnamefont
  {Albert}}, \bibinfo {author} {\bibfnamefont {B.}~\bibnamefont {Bradlyn}},
  \bibinfo {author} {\bibfnamefont {M.}~\bibnamefont {Fraas}}, \ and\ \bibinfo
  {author} {\bibfnamefont {L.}~\bibnamefont {Jiang}},\ }\bibfield  {title}
  {\emph {\bibinfo {title} {Geometry and response of lindbladians},\ }}\href
  {\doibase 10.1103/PhysRevX.6.041031} {\bibfield  {journal} {\bibinfo
  {journal} {Phys. Rev. X}\ }\textbf {\bibinfo {volume} {6}},\ \bibinfo {pages}
  {041031} (\bibinfo {year} {2016})}\BibitemShut {NoStop}%
\bibitem [{\citenamefont {Minganti}\ \emph {et~al.}(2018)\citenamefont
  {Minganti}, \citenamefont {Biella}, \citenamefont {Bartolo},\ and\
  \citenamefont {Ciuti}}]{PhysRevA.98.042118}%
  \BibitemOpen
  \bibfield  {author} {\bibinfo {author} {\bibfnamefont {F.}~\bibnamefont
  {Minganti}}, \bibinfo {author} {\bibfnamefont {A.}~\bibnamefont {Biella}},
  \bibinfo {author} {\bibfnamefont {N.}~\bibnamefont {Bartolo}}, \ and\
  \bibinfo {author} {\bibfnamefont {C.}~\bibnamefont {Ciuti}},\ }\bibfield
  {title} {\emph {\bibinfo {title} {{Spectral theory of Liouvillians for
  dissipative phase transitions}},\ }}\href {\doibase
  10.1103/PhysRevA.98.042118} {\bibfield  {journal} {\bibinfo  {journal} {Phys.
  Rev. A}\ }\textbf {\bibinfo {volume} {98}},\ \bibinfo {pages} {042118}
  (\bibinfo {year} {2018})}\BibitemShut {NoStop}%
\bibitem [{\citenamefont {Minganti}\ \emph {et~al.}(2019)\citenamefont
  {Minganti}, \citenamefont {Miranowicz}, \citenamefont {Chhajlany},\ and\
  \citenamefont {Nori}}]{PhysRevA.100.062131}%
  \BibitemOpen
  \bibfield  {author} {\bibinfo {author} {\bibfnamefont {F.}~\bibnamefont
  {Minganti}}, \bibinfo {author} {\bibfnamefont {A.}~\bibnamefont
  {Miranowicz}}, \bibinfo {author} {\bibfnamefont {R.~W.}\ \bibnamefont
  {Chhajlany}}, \ and\ \bibinfo {author} {\bibfnamefont {F.}~\bibnamefont
  {Nori}},\ }\bibfield  {title} {\emph {\bibinfo {title} {{Quantum exceptional
  points of non-Hermitian Hamiltonians and Liouvillians: The effects of quantum
  jumps}},\ }}\href {\doibase 10.1103/PhysRevA.100.062131} {\bibfield
  {journal} {\bibinfo  {journal} {Phys. Rev. A}\ }\textbf {\bibinfo {volume}
  {100}},\ \bibinfo {pages} {062131} (\bibinfo {year} {2019})}\BibitemShut
  {NoStop}%
\bibitem [{\citenamefont {Lieu}\ \emph {et~al.}(2020)\citenamefont {Lieu},
  \citenamefont {McGinley},\ and\ \citenamefont
  {Cooper}}]{PhysRevLett.124.040401}%
  \BibitemOpen
  \bibfield  {author} {\bibinfo {author} {\bibfnamefont {S.}~\bibnamefont
  {Lieu}}, \bibinfo {author} {\bibfnamefont {M.}~\bibnamefont {McGinley}}, \
  and\ \bibinfo {author} {\bibfnamefont {N.~R.}\ \bibnamefont {Cooper}},\
  }\bibfield  {title} {\emph {\bibinfo {title} {{Tenfold Way for Quadratic
  Lindbladians}},\ }}\href {\doibase 10.1103/PhysRevLett.124.040401} {\bibfield
   {journal} {\bibinfo  {journal} {Phys. Rev. Lett.}\ }\textbf {\bibinfo
  {volume} {124}},\ \bibinfo {pages} {040401} (\bibinfo {year}
  {2020})}\BibitemShut {NoStop}%
\bibitem [{\citenamefont {Deng}\ \emph {et~al.}(2021)\citenamefont {Deng},
  \citenamefont {Pan}, \citenamefont {Chen},\ and\ \citenamefont
  {Zhai}}]{PhysRevLett.127.086801}%
  \BibitemOpen
  \bibfield  {author} {\bibinfo {author} {\bibfnamefont {T.-S.}\ \bibnamefont
  {Deng}}, \bibinfo {author} {\bibfnamefont {L.}~\bibnamefont {Pan}}, \bibinfo
  {author} {\bibfnamefont {Y.}~\bibnamefont {Chen}}, \ and\ \bibinfo {author}
  {\bibfnamefont {H.}~\bibnamefont {Zhai}},\ }\bibfield  {title} {\emph
  {\bibinfo {title} {{Stability of Time-Reversal Symmetry Protected Topological
  Phases}},\ }}\href {\doibase 10.1103/PhysRevLett.127.086801} {\bibfield
  {journal} {\bibinfo  {journal} {Phys. Rev. Lett.}\ }\textbf {\bibinfo
  {volume} {127}},\ \bibinfo {pages} {086801} (\bibinfo {year}
  {2021})}\BibitemShut {NoStop}%
\end{thebibliography}
\end{document}